\documentclass[prd, aps, nofootinbib, preprint, letterpaper]{revtex4-2}

\usepackage{mathtools}
\usepackage{color}
\definecolor{darkred}{RGB}{196,0,0}

\usepackage{amsmath, amssymb, amsfonts}
\usepackage[hidelinks]{hyperref}
\usepackage{ulem}
\usepackage{setspace}
    
\newcommand{\be}{\begin{equation}}
\newcommand{\ee}{\end{equation}}
\newcommand{\ba}{\begin{eqnarray}}
\newcommand{\ea}{\end{eqnarray}}

\newcommand{\uh}{\hat \nu}

\newcommand{\oh}{\hat\omega}

\begin{document}

\title {Energy loss of a heavy quark in a collisional quark-gluon plasma}

\author{Mingda Cai}
\author{Yun Guo$^*$}
\affiliation{Department of Physics, Guangxi Normal University, Guilin, 541004, China}
\affiliation{Guangxi Key Laboratory of Nuclear Physics and Technology, Guilin, 541004, China}

\renewcommand{\thefootnote}{\fnsymbol{footnote}}
\footnotetext[1]{Contact author: yunguo@mailbox.gxnu.edu.cn}
\renewcommand{\thefootnote}{\arabic{footnote}}

\begin{abstract}
We extend our previous work on the energy loss of a heavy fermion in a QED plasma to the quark-gluon plasma, using the same Bhatnagar-Gross–Krook collisional kernel. The calculation is carried out with a theoretical method where the hard-thermal-loop resummed gluon propagator is used for arbitrary momentum transfer in the scattering processes. Encoding the collision effect in a self-consistent manner, the resummed gluon propagator regulates the infrared divergence in the scattering amplitude without introducing an artificial cutoff for the transferred momenta and makes the analysis on the hard and soft processes in a unified framework. 
To place our computation on a more solid foundation, we also explicitly demonstrate the gauge independence of the interaction rate as well as the elimination of unphysical gluon polarizations by the ghost field under the use of the resummed gluon propagator. 
In addition, with our complete QCD calculation by including both quark-quark and quark-gluon scatterings, we provide a quantitatively reliable result on the collisional energy loss of a heavy quark where the new contribution from quark-gluon scatterings accounts for a larger portion of the total energy loss. In general, collisions between the thermal partons result in an increased energy loss which becomes more pronounced with increasing gauge coupling. Considering a typical coupling constant in QCD, $\alpha_s = 0.3$, the energy loss increases $\sim 8 \%$ at large incident velocities as compared to the collisionless limit. Such a collision-induced correction is still moderate although it is slightly suppressed as compared to our previous estimate based on a QED calculation using couplings consistent with those expected to be generated in the quark-gluon plasma.  
\end{abstract}

\maketitle

\newpage

\section{Introduction}

Heavy quarks can serve as ideal probes of the quark-gluon plasma (QGP), a primordial state of matter existed in the early Universe. Produced predominantly through hard processes at a very early stage in ultrarelativistic heavy-ion collisions, charm and bottom quarks propagate through the entire evolution of the QGP, interacting with its constituents and losing energy. The mechanism of the energy loss is deeply connected to the fundamental properties of the QGP. Therefore, the study of heavy-quark energy loss has drawn a lot of attention during the past decades~\cite{Bjorken:1982tu,Gyulassy:1990bh,Thoma:1990fm,Mrowczynski:1991da,Braaten:1991we,Romatschke:2003vc,Romatschke:2004au,Peigne:2007sd,Peigne:2008nd,Carignano:2021mrn,Du:2024riq,Lin:2013efa}. 

For massive quarks, small-angle gluon radiation is suppressed due to the dead-cone effect~\cite{Dokshitzer:2001zm,Zhang:2003wk,Mustafa:2004dr,Wicks:2005gt}. Therefore, besides the radiated energy loss from inelastic gluon radiation~\cite{Gyulassy:1993hr,Baier:1996kr,Zakharov:1996fv,Gyulassy:2000fs,Wiedemann:2000za,Wang:2001ifa,Arnold:2001ms,Arnold:2002ja,Djordjevic:2003zk}, collisional energy loss induced by elastic scatterings also plays an important role to interpret the observed jet suppression at both the Relativistic Heavy Ion Collider and LHC~\cite{Qin:2007rn,Schenke:2009ik,Schenke:2009gb}. The collisional energy loss of a heavy fermion in high-temperature plasmas has been extensively studied in previous literature, where a commonly used technique is to introduce a cutoff scale $q^*$ for the transferred momenta to separate the hard and soft scatterings~\cite{Braaten:1991dd}. The hard contribution to the energy loss is obtained by a diagrammatic calculation with bare propagators, while infrared divergence in the soft scatterings can be regulated by using the hard-thermal-loop (HTL) resummed propagator. However, the resulting energy loss would manifestly depend on this artificial separation scale $q^*$. It was found that such a $q^*$ dependence could be exactly canceled between the hard and soft contributions in the small-coupling limit where $g T\ll q^* \ll T$ was satisfied~\cite{Braaten:1991we}. On the other hand, even at currently achievable
collision energies in heavy-ion experiments, considering a moderate coupling constant is necessary. In Refs.~\cite{Romatschke:2003vc,Romatschke:2004au}, a numerical evaluation showed that there was a residual dependence on the cutoff at any finite gauge coupling. Given the typical values of the QCD coupling constant $\alpha_s=0.3$, the $q^*$ dependence of the collisional energy loss is actually non-negligible.

There is an alternative way to compute the collisional energy loss without introducing the above-mentioned cutoff scale. In this case, energy loss is also computed based on the tree-level Feynman diagrams, but using a HTL resummed gluon propagator in the $t$-channel scatterings for arbitrary momentum transfers. It was proposed in Ref.~\cite{Braaten:1991we} and first applied in Ref.~\cite{Djordjevic:2006tw} to study the finite-size effect on the collisional energy loss. We also adopted this method to compute the energy loss of a heavy muon in a collisional QED plasma~\cite{Guo:2024mgh}. In all the mentioned works, explicit calculations have been carried out only for quark-quark scattering. We will include the contribution from quark-gluon scatterings in the current work, which is important to get a quantitatively reliable result on the collisional energy loss of a heavy quark. In addition, the validity of this theoretical method, especially using a resummed propagator instead of a bare propagator, still needs to be addressed, because the prescription with a resummed propagator cannot be trivially derived from the Feynman rules of thermal field theory. To do so, we will demonstrate the gauge independence of the full QCD calculation as well as the elimination of unphysical gluon polarizations by ghost field under the use of the resummed propagator.

It is helpful to mention that incorporating finite-size effects and dynamical scattering centers into the study on elastic and radiative energy loss was considered in Refs.~\cite{Djordjevic:2006tw,Djordjevic:2007at}, where the calculations also rely on the use of the HTL resummed propagator. Similarly, with the effective dynamical quasiparticle model, elastic and inelastic reactions between the heavy quark and medium partons have been evaluated by using a model-based ``dressed" propagator instead of the bare one~\cite{Grishmanovskii:2025mnc}. On the other hand, as shown in Refs.~\cite{Faraday:2024gzx,Faraday:2024zfj}, depending on the theoretical method adopted in the calculation, the transition between vacuum and HTL propagators could introduce significant uncertainties in predicting the jet quenching observables. In this work, in order to show the discrepancy among the different theoretical methods, a systematic comparison of the corresponding energy loss will be presented.

Another focus in this work is the collision-induced corrections to the heavy-quark energy loss. Given a number-conserving Bhatnagar-Gross-Krook
(BGK) collisional kernel, these corrections can be taken into account in a self-consistent manner by using a collisionally modified resummed gluon propagator. Consequently, the above theoretical method can be straightforwardly generalized to study the energy loss in a collisional plasma. In this approach, collisions between medium partons affect all the scattering processes regardless of the magnitude of the transferred momenta, although the collision effect becomes less accentuated for hard processes. Based on plasma-physics techniques as used in Ref.~\cite{Thoma:1990fm}, previous works~\cite{Han:2017nfz,Shi:2018aeb,YousufJamal:2019pen,Jamal:2020emj} investigated the same problem by considering only soft contributions. However, this is not sufficient at moderate gauge couplings, as the soft contributions are not dominated anymore. In fact, collision effect also becomes irrelevant in the small-coupling limit due to the extremely small collision rate. With our complete QCD calculation, the collision-induced corrections to the energy loss can be analyzed in a systematical way.

The structure of our paper is as follows. In Sec.~\ref{re}, we present a full QCD calculation of the interaction rate for scatterings between the heavy quark and medium partons, from which the collisional energy loss can be evaluated. To assess the dependence of our results on the calculational scheme, we also provide numerical comparisons of the energy loss obtained by different theoretical methods. 
In Sec.~\ref{gauge}, to further justify the theoretical method used in this work, we explicitly show the gauge independence of the interaction rate under the use of the HTL resummed gluon propagator. In addition, an expected behavior that the elimination of unphysical gluon polarizations by ghost contribution is also verified under the HTL approximation. In Sec.~\ref{elwithcolli}, using the collisionally modified gluon propagator, we numerically evaluate the heavy-quark energy loss in a collisional QGP which is described by the BGK collisional kernel. With phenomenological values for the collision rate, we present our quantitative results on the energy loss of a charm or bottom quark with emphasis on the enhancement caused by the collision effect. Our complete QCD calculation amends the previous estimate on the collision-induced corrections based on a QED calculation. Finally, conclusions and outlook are presented in Sec.~\ref{con}.

\section{Collisional energy loss of a heavy quark in the Quark-Gluon plasma}\label{re}

When a heavy quark with mass $M$ and four-momentum $P=(E,{\bf p})$ traverses through the QGP at a temperature $T$, it may lose energy by interacting with the constituent particles in the hot medium. The rate of energy loss per distance traveled is given by
\be\label{def}
-\frac{d E}{d x}=\frac{1}{v}\int_M^\infty
dE^\prime (E-E') \frac{d \Gamma}{d E'}\, ,
\ee
where the velocity of the incident heavy quark is given by ${\bf v}={\bf p}/E$. The interaction rate $\Gamma(E)$ can be expressed in terms of the squared matrix element for the corresponding scattering processes, weighted by the thermal distribution functions of the medium partons. Explicitly, we have
\ba
\Gamma (E)&=&\frac{1}{2E} \int \frac{d^3 {\bf{p}}'}{(2\pi)^3 2{E}'}\int \frac{d^3 {\bf k}}{(2\pi)^3 2k} n_{B/F}(k)\int \frac{d^3 {\bf {k}}'}{(2\pi)^3 2{k}'}[1 \pm n_{B/F}({k}')]\,\nonumber \\
&& \hspace{2cm}\times (2\pi)^4\delta^4(P+K-{P}'-{K}')\,\bigg(\frac{1}{6}\sum|\mathcal{M}|^2\bigg)\, .
\label{interaction_rate}
\ea
In the above equation, the sum runs over colors and spins of the incoming and outgoing partons. The four-momentum of the thermal light quark or gluon that scatters off the incident heavy quark is denoted by $K=(k, {\bf k})$, and the primed variables correspond to the energies or momenta of the outgoing heavy-quark and medium partons. For quark-quark scattering, the Pauli-blocking factor $1- n_{F}(k^\prime)$ is included in evaluating the interaction rate where the Fermi-Dirac distribution $n_{F}(k)=(e^{k/T}+1)^{-1}$, and for quark-gluon scattering, one needs to use the Bose-enhanced factor $1+ n_{B}(k^\prime)$ with the Bose-Einstein distribution $n_{B}(k)=(e^{k/T}-1)^{-1}$. According to Eq.~(\ref{def}), the energy loss $-d E/d x$ can be computed by inserting $(E-E')/v\equiv \omega/v$ into the integrals in Eq.~(\ref{interaction_rate}).

\begin{figure}[htbp]
\begin{center}
\includegraphics[width=0.8\linewidth]{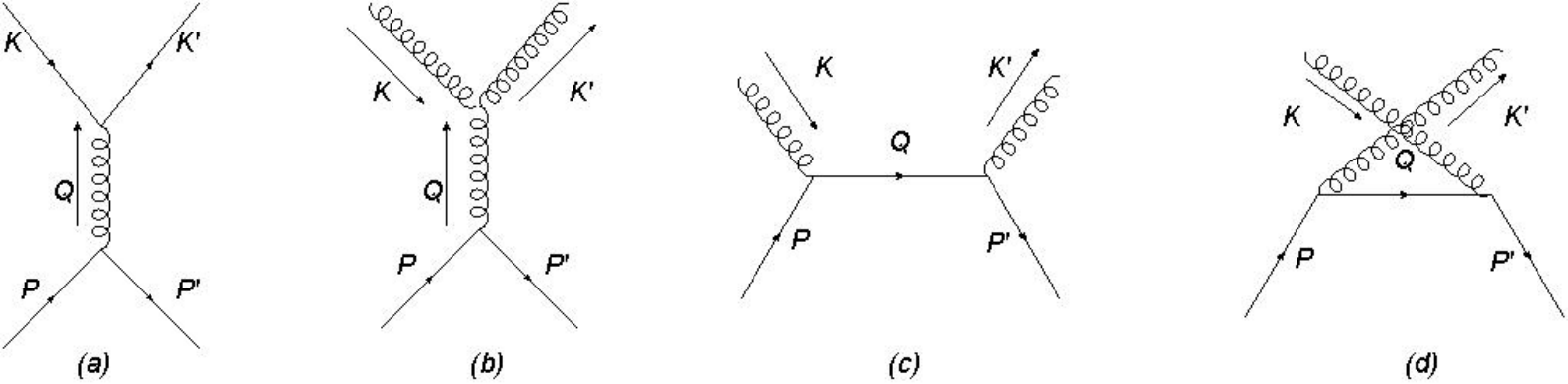}
\caption{The tree-level Feynman diagrams for the elastic scatterings $Q q \rightarrow Q q$ and $Q g\rightarrow Q g$.}
\label{lod}
\end{center}
\end{figure}

When scattering processes between the heavy quark and thermal partons involve hard momentum transfer $\sim T$, we can consider only the tree-level Feynman diagrams as shown in Fig.~\ref{lod}. On the other hand, for $t$-channel diagrams as given by Figs.~\ref{lod}(a) and \ref{lod}(b), self-energy insertion into the bare gluon propagator becomes essential for soft momentum transfer $\sim gT$, which plays an important role to regulate the infrared divergence.\footnote{We assume the mass of the incident heavy quark is very large $M_Q\gg T$ and the self-energy correction to the bare quark propagator appearing in Figs.~\ref{lod}(c) and \ref{lod}(d) is negligible, as the heavy quark is not thermalized.} In order to deal with the hard and soft processes in a unified framework, following the same method as used in our previous work~\cite{Guo:2024mgh}, we will adopt an effective gluon propagator, i.e., the HTL resummed propagator, to compute the matrix element for the $t$-channel contributions for arbitrary momentum transfer. The HTL resummed gluon propagator $D^{\mu\nu}(Q)$ reads
\begin{equation}\label{repro}
- i D^{\mu\nu}(Q) = {\cal A}(Q) g^{\mu\nu} + {\cal B}(Q) M^\mu M^\nu \, ,
\end{equation}
where the four-momentum of the exchanged gluon is denoted by $Q=K^\prime-K=(\omega, {\bf q})$ and  $M^\mu = (1, 0, 0, 0)$ is the four-velocity of the thermal rest frame. The scalar coefficients ${\cal A}$ and ${\cal B}$ are given by
\begin{equation}
 {\cal A}(Q) = -\frac{1}{Q^2 - \Pi_T(\hat{\omega})}\, 
\quad{\rm and}\quad
{\cal B}(Q) = \frac{1 - \hat{\omega}^2}{Q^2 - \Pi_T(\hat{\omega})} + \frac{1}{q^2 - \Pi_L(\hat{\omega})}\, ,
\end{equation}
with the dimensionless variable $\hat{\omega} \equiv \omega/q$. The transverse and longitudinal HTL gluon self-energies take the following forms:
\ba\label{pi}
\Pi_T(\hat{\omega}) &=& \frac{m_D^2}{2} \hat{\omega}^2 \left[1 - \frac{\hat{\omega}^2 - 1}{2 \hat{\omega}} \ln \left(\frac{\hat{\omega} + 1 +i \epsilon}{\hat{\omega} - 1 +i \epsilon}\right)\right]\, , \nonumber \\
\Pi_L(\hat{\omega}) &=& m_D^2 \left[-1 + \frac{\hat{\omega}}{2} \ln \left(\frac{\hat{\omega} + 1+i \epsilon}{\hat{\omega} - 1+i \epsilon}\right)\right]\, ,
\ea
and the QCD Debye screening mass is 
\begin{equation}
m_D^2 = \frac{g^2 T^2}{3} \left(N_c + \frac{N_f}{2}\right)\, ,
\end{equation}
where $N_c=3$ and $N_f$ denotes the number of flavors for the thermal quarks. We choose $N_f=2$ throughout this work.

We should mention that, in general linear gauges, $D^{\mu\nu}(Q)$ also has terms proportional to $Q^\mu M^\nu+Q^\nu M^\mu$ and $Q^\mu Q^\nu$ which are omitted in Eq.~(\ref{repro}). The explicit forms of these terms depend on the specific gauges. However, as we will show in Sec.~\ref{gain}, the gauge-dependent terms in the resummed gluon propagator have no contributions to the gauge-independent $\Gamma(E)$. Therefore, in the following calculations, we keep only the gauge-independent terms proportional to $g^{\mu\nu}$ and $M^\mu M^\nu$ as given in Eq.~(\ref{repro}).

The squared matrix element $|{\cal M}|^{2}$ obtained by using $D^{\mu\nu}(Q)$ in Eq.~(\ref{repro}) takes a very complicated form. It can be simplified by assuming $M_Q\gg T$, $p \gg T$, and $E\ll M_Q^2/T$; the corresponding results for quark-quark\footnote{There is an extra factor of $2$ introduced in Eq.~(\ref{mqq}) to account for the contribution from scattering between the heavy quark and an antiquark.} and quark-gluon scattering (only $t$-channel contribution) are given by
\ba\label{mqq}
\frac{1}{6}\sum|{\cal M}|_{Qq}^2&=&\frac{64}{3} N_f g^4\Big\{ |\Delta_L(Q)|^2 E^{2} (\mathbf{k} \cdot \mathbf{k'} + k k') +2 {\rm Re} \big(\Delta_L(Q) \Delta_T(Q)^*\big)\nonumber \\
&\times&E\big[k\big({\bf p} \cdot{\bf k}^\prime - ({\bf p} \cdot {\hat {\bf q}})( {\bf k}^\prime \cdot {\hat {\bf q}}) \big)+k^\prime \big({\bf p} \cdot{\bf k} - ({\bf p} \cdot {\hat {\bf q}})( {\bf k}\cdot {\hat {\bf q}}) \big)\big]\nonumber \\
&+& |\Delta_T(Q)|^2\big[2\big({\bf p} \cdot{\bf k} - ({\bf p} \cdot {\hat {\bf q}})( {\bf k} \cdot {\hat {\bf q}})\big)\big({\bf p} \cdot{\bf k}^\prime - ({\bf p} \cdot {\hat {\bf q}})( {\bf k}^\prime \cdot {\hat {\bf q}})\big)
\nonumber \\
&+&(k k^\prime-{\bf k} \cdot{\bf k}^\prime)\big(p^2 - ({\bf p} \cdot {\hat {\bf q}})( {\bf p} \cdot {\hat {\bf q}}) \big)\big]\Big\}\,
\ea
and
\ba\label{mqg}
\frac{1}{6}\sum|{\cal M}|_{Qg}^2&=&128 g^4\Big\{ |\Delta_L(Q)|^2 E^{2} \left[\mathbf{k} \cdot \mathbf{k'}  -(k - k')^{2}/4\right ]+ {\rm Re} \big(\Delta_L(Q) \Delta_T(Q)^*\big)\nonumber \\
&\times& E\big[k\big({\bf p} \cdot{\bf k}^\prime - ({\bf p} \cdot {\hat {\bf q}})( {\bf k}^\prime \cdot {\hat {\bf q}}) \big)+k^\prime \big({\bf p} \cdot{\bf k} - ({\bf p} \cdot {\hat {\bf q}})( {\bf k}\cdot {\hat {\bf q}}) \big)\big]\nonumber \\
&+& |\Delta_T(Q)|^2\big[\big({\bf p} \cdot{\bf k} - ({\bf p} \cdot {\hat {\bf q}})( {\bf k} \cdot {\hat {\bf q}})\big)\big({\bf p} \cdot{\bf k}^\prime - ({\bf p} \cdot {\hat {\bf q}})( {\bf k}^\prime \cdot {\hat {\bf q}})\big)
\nonumber \\
&+&(k k^\prime-{\bf k} \cdot{\bf k}^\prime)\big(p^2 - ({\bf p} \cdot {\hat {\bf q}})( {\bf p} \cdot {\hat {\bf q}}) \big)\big]\Big\}\,, 
\ea
respectively. In the above equations, we define ${\hat {\bf q}}={\bf q}/q$. In fact, Eq.~(\ref{mqq}) can be easily obtained from the corresponding QED result multiplied by a trivial color factor, while Eq.~(\ref{mqg}) is new and, in general, there is no simple proportionality between the above two equations. 
In Ref.~\cite{Djordjevic:2006tw}, based on the quark-quark scattering, a full QCD result for the collisional energy loss was obtained by assuming a simple proportionality between the two squared matrix elements. We point out that this is valid only for soft momentum exchange where one can approximate ${\bf k}\approx {\bf k}^\prime$. Under this assumption, the following simple relation can be obtained:
\be\label{ratiom}
\sum|{\cal M}|_{Qq}^2 \approx \frac{N_f}{3} \sum|{\cal M}|_{Qg}^2\, \quad\quad{\rm for}\quad\quad {\bf k}\approx {\bf k}^\prime\, .
\ee

For an isotropic medium as considered in this work, the energy loss does not depend on the incident direction of the heavy quark which indicates the following identity:
\be\label{av}
-\frac{d E}{d x}=\int \frac{d\Omega_{{\bf v}}}{4\pi}\,\left(-\frac{d E}{d x}\right)\, .
\ee
Consequently, after averaging over all the directions of ${\bf v}$, the collisional energy loss of the heavy quark can be expressed as
\ba\label{elqq}
-\bigg(\frac{dE}{dx}\bigg)_{Qq}&=&\frac{8\alpha_s^2 N_f}{3\pi v^2}\int_{0}^{\infty} d k\,  n_F(k) \left(\int_{0}^{2k/(1+v)} dq\,q^2 \int_{-v q}^{v q} d\omega \, \omega+\int_{2k/(1+v)}^{2k/(1-v)} dq\,q^2 \int_{q-2k}^{v q} d\omega \, \omega\right) \nonumber\\& \times& \bigg[ |\Delta_L(Q)|^2 f_1(k,q,\omega)+(1-{\hat{\omega}^2})(v^2-{\hat{\omega}^2}) |\Delta_T(Q)|^2 f_2(k,q,\omega)\bigg]\, 
\ea
and
\ba\label{elqg}
-\bigg(\frac{dE}{dx}\bigg)_{Qg} &=& \frac{8 \alpha _{s}^{2}  }{\pi v^{2}} \int_{0}^{\infty} dk\, n_{B}(k) \left( \int_{0}^{2k/(1+v)} dq\, q^{2} \int_{-vq}^{vq} d\omega\,\omega 
+ \int_{2k/(1+v)}^{2k/(1-v)} dq\, q^{2} \int_{q-2k}^{vq} d\omega\,\omega \right)  \nonumber \\
&\times& \bigg[ \left|\Delta_{L}(Q)\right|^{2} \big(f_{1}(k,q,\omega)-\frac{1}{2}\big) +(1-{\hat{\omega}^2})(v^2-{\hat{\omega}^2}) |\Delta_T(Q)|^2  \big(f_2(k,q,\omega)+\frac{1}{2}\big) \bigg]\, , \nonumber \\
\ea
where
\be\label{protl}
\Delta_T(Q)=\frac{1}{Q^2-\Pi_T(\oh)}\,,\quad \quad
\Delta_L(Q)=\frac{1}{q^2-\Pi_L(\oh)}\, ,
\ee
and
\be
f_1(k,q,\omega)= 2 k\frac{\omega+k}{q^2}+\frac{{\hat{\omega}^2}-1}{2}\,,\quad\quad\quad
f_2(k,q,\omega)= k\frac{\omega+k}{q^2}+\frac{{\hat{\omega}^2}+1}{4}\, .
\ee
Notice that the cross terms $\sim {\rm Re} \big(\Delta_L(Q) \Delta_T(Q)^*\big)$ in Eqs.~(\ref{mqq}) and (\ref{mqg}) do not contribute to the energy loss after performing the average as given in Eq.~(\ref{av}).

To get the above results, we have changed the integral variables as follows\footnote{The integral over ${\bf p}^{\prime}$ can be carried out by using the delta function in Eq.~(\ref{interaction_rate}).}:
\be\label{changeva}
\frac{1}{2(2\pi)^2}\int \frac{d^3 \mathbf{k}}{k} \int \frac{d^3\mathbf{k}'}{k^\prime}  \to  \int_{0}^{\infty} dk \left( \int_{0}^{2k/(1+v)}q dq \int_{-vq}^{vq} d\omega 
+ \int_{2k/(1+v)}^{2k/(1-v)}q dq\int_{q-2k}^{vq} d\omega \right) 
\, 
\ee
and used an approximated energy conservation $\delta(w-{\bf v}\cdot{\bf q})$ which is consistent with those assumptions for simplifying the squared matrix element. For quark-gluon scattering, contributions from ghost field have been included to cancel the unphysical gluon polarizations. We should also mention that our results do not apply to the special case $v\rightarrow 1$, where one instead assumes $E\gg M_Q^2/T$. The collisional energy loss in the ultrarelativistic limit has been studied in Ref.~\cite{Peigne:2007sd}.
 
At this point, it is worthwhile to mention a different method to study the collisional energy loss which has been widely used in previous works. As first proposed in Ref.~\cite{Braaten:1991dd}, by introducing a momentum cutoff $q^*$, contributions from hard and soft scatterings are computed separately with a theta function $\theta(q-q^*)$ and $\theta(q^*-q)$, respectively. Clearly, the appearance of the theta functions imposes a constraint on the possible values of the transferred momenta. The main purpose of such a treatment is to eliminate the infrared divergence caused by using the bare gluon propagator. On the other hand, there is no need to consider a cutoff $q^*$ in Eqs.~(\ref{elqq}) and (\ref{elqg}), since the divergence is regulated by the resummed gluon propagator. Despite these differences, we can show that our results agree with those obtained in previous works in the hard and soft limit. To make a comparison, we formally define the hard and soft contributions by modifying the integral intervals in Eqs.~(\ref{elqq}) and (\ref{elqg}). For hard momentum transfer, the integrals read 
\ba\label{changevah}
\frac{1}{2(2\pi)^2}\int \frac{d^3 \mathbf{k}}{k} \int \frac{d^3\mathbf{k}'}{k^\prime} \theta(q-q^*) &\to& \int_{\frac{1+v}{2}{q^*}}^{\infty} d k\int_{ q^*}^{\frac{2k}{1+v}} q dq \int_{-v q}^{v q} d\omega 
+\int_{\frac{1+v}{2}{ q^*}}^{\infty} d k\int_{\frac{2k}{1+v}}^{\frac{2k}{1-v}}q dq \int_{q-2k}^{v q} d\omega 
\nonumber \\&+&\int_{\frac{1-v}{2}{q^*}}^{\frac{1+v}{2}{q^*}} d k\int_{q^*}^{\frac{2k}{1-v}} q dq \int_{q-2k}^{v q} d\omega\,.
\ea
For soft momentum transfer, we should adopt the following integrals:
\ba\label{changevas}
\frac{1}{2(2\pi)^2}\int \frac{d^3 \mathbf{k}}{k} \int \frac{d^3\mathbf{k}'}{k^\prime} \theta(q^*-q) &\to& \int_{\frac{1+v}{2}{q^*}}^{\infty} d k\int_{0}^{ q^*} q dq \int_{-v q}^{v q} d\omega + \int_0^{\frac{1+v}{2}{q^*}} d k\int_{0}^{\frac{2k}{1+v}} q dq \int_{-v q}^{v q} d\omega 
\nonumber \\&&\hspace{-1cm} +\int_0^{\frac{1+v}{2}{ q^*}} d k\int_{\frac{2k}{1+v}}^{\frac{2k}{1-v}}q dq \int_{q-2k}^{v q} d\omega 
+\int_{\frac{1-v}{2}{q^*}}^{\frac{1+v}{2}{q^*}} d k\int^{q^*}_{\frac{2k}{1+v}} q dq \int_{q-2k}^{v q} d\omega\,. \nonumber \\
\ea
By definition, the sum of Eqs.~(\ref{changevah}) and (\ref{changevas}) leads to Eq.~(\ref{changeva}); therefore, the total energy loss has no dependence on the cutoff $q^*$ in our approach.

To reproduce the energy loss induced by hard scatterings as obtained in Ref.~\cite{Romatschke:2004au} [denoted as the Romatschke-Strickland (RS) result], we need only to replace $\Delta_T(Q)\rightarrow 1/Q^2$ and $\Delta_L(Q)\rightarrow 1/q^2$ in Eqs.~(\ref{elqq}) and (\ref{elqg}) and adopt the integrals in Eq.~(\ref{changevah}). This is actually trivial, because these replacements correspond to the use of a bare gluon propagator which is just the strategy used in Ref.~\cite{Romatschke:2004au}. On the other hand, to reproduce the RS result for soft processes, we assume $q^*\ll T \sim k$, and the integrals in Eq.~(\ref{changevas}) are simplified into
\be\label{changeva4}
\frac{1}{2(2\pi)^2}\int \frac{d^3 \mathbf{k}}{k} \int \frac{d^3\mathbf{k}'}{k^\prime}\theta(q^*-q)  \to  \int_{0}^{\infty} dk \int_{0}^{q^*}q dq \int_{-vq}^{vq} d\omega 
\, .
\ee
Taking into account the condition $\omega, q\ll k$ for soft processes, we have $f_1(k,q,\omega)\approx 2f_2(k,q,\omega)\approx 2(k^2+k \omega)/q^2\gg 1$. Here, terms $\sim k^2/q^2$ do not contribute because of the integration over $\omega$. Therefore, with the nonvanishing leading-order terms $\sim k \omega/q^2$, the integral over $k$ can be performed analytically, and we arrive at
\be\label{elqqs}
-\bigg(\frac{dE}{dx}\bigg)_{Qq}^{\rm soft}=\frac{N_f}{6}g^2 T^2 \frac{g^2 }{6\pi v^2}\int_{0}^{q^*} dq\, \int_{-v q}^{v q} d\omega \, \omega^2  \bigg[|\Delta_L(Q)|^2 +\frac{1-\hat{\omega}^2}{2}(v^2-\hat{\omega}^2) |\Delta_T(Q)|^2 \bigg]\, 
\ee
and
\be\label{elqgs}
-\bigg(\frac{dE}{dx}\bigg)_{Qg}^{\rm soft} =g^2 T^2 \frac{g^{2}}{6\pi v^{2}} \int_{0}^{q^*} dq\, \int_{-vq}^{vq} d\omega\,\omega^2 
 \bigg[\left|\Delta_{L}(Q)\right|^{2}  +\frac{1-\hat{\omega}^2}{2}(v^2-\hat{\omega}^2) |\Delta_T(Q)|^2   \bigg]\, ,
\ee
which show that there is a simple proportionality existing between Eqs.~(\ref{elqqs}) and~(\ref{elqgs}):
\be
-\bigg(\frac{dE}{dx}\bigg)_{Qq}^{\rm soft} =-\frac{N_f
}{6}\bigg(\frac{dE}{dx}\bigg)_{Qg}^{\rm soft}\,.
\ee
Adding up the contributions from quark-quark and quark-gluon scatterings, it is clear to see that, due to the just right proportional coefficient $N_f/6$ in the above equation, the heavy-quark energy loss from soft processes is proportional to the squared Debye mass $m_D^2=g^2 T^2(1+N_f/6)$. This is a highly nontrivial conclusion based on our approach to computing the collisional energy loss. Obviously, Eq.~(\ref{ratiom}) is very essential to guarantee this conclusion. Given the proportional coefficient $N_f/3$ in Eq.~(\ref{ratiom}), another necessary condition for reaching the conclusion is that the ratio between the two integrals over $k$ involving $n_B(k)$, corresponding to quark-gluon scattering, and $n_F(k)$, corresponding to quark-quark scattering, should be exactly equal to 2. It actually holds because these two integrals are just given by $\int k n_B(k) dk$ and $\int k n_F(k) dk$ in our calculation.

Collisional energy loss induced by soft scatterings as given by Eqs.~(\ref{elqqs}) and (\ref{elqgs}) is in agreement with the RS result which was obtained by computing the chromoelectric field induced by a classical charge moving at a fixed velocity. Therefore, the above discussions have shown that our method with the HTL resummed gluon propagator leads to the same energy loss as the RS result. We should emphasize that the agreement holds only in the hard and soft limit. To the best our knowledge, the exact match in the soft limit is first demonstrated for the full QCD case in the current work. 

\begin{figure}[htbp]
\begin{center}
\includegraphics[width=0.49\linewidth]{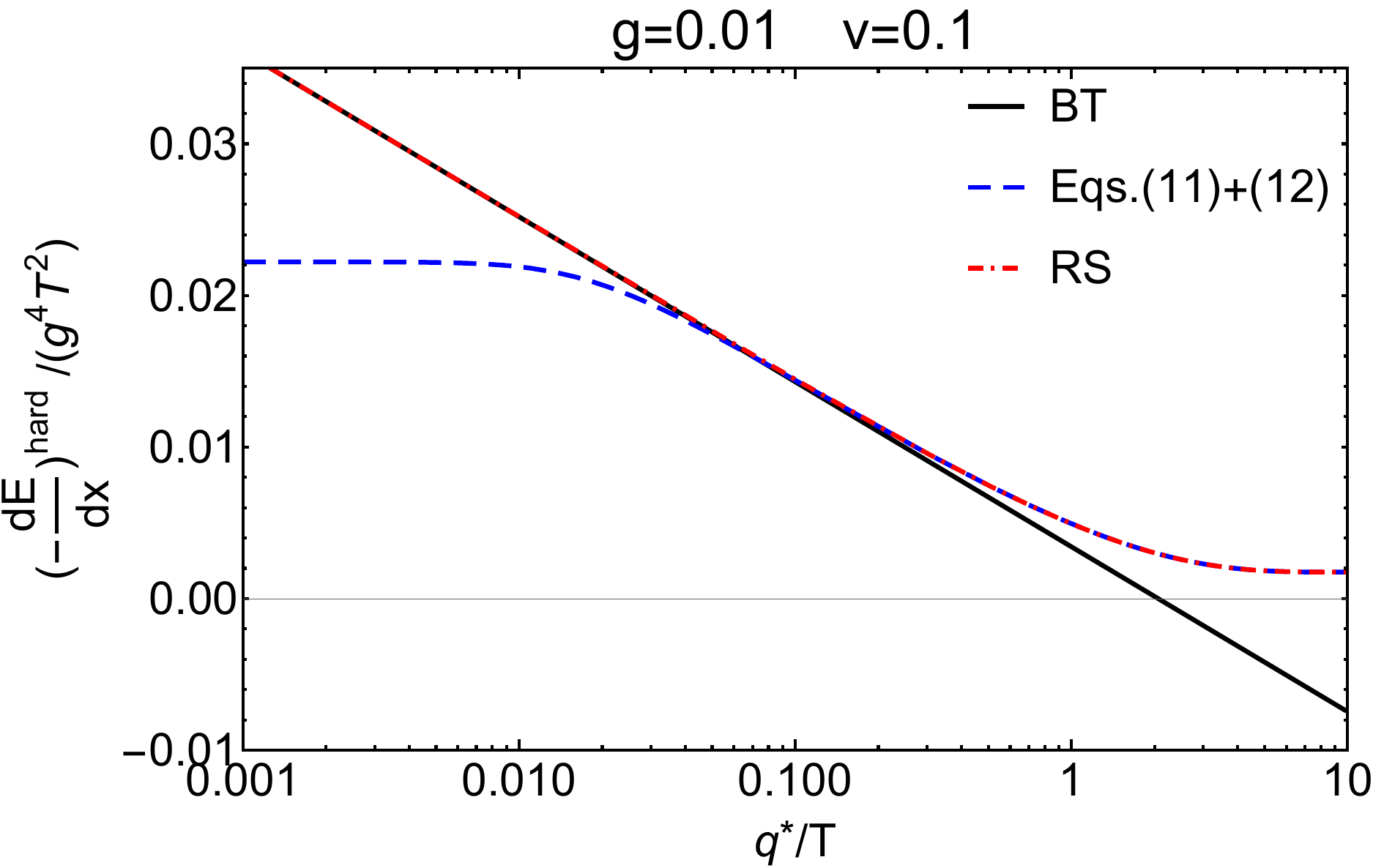}
\includegraphics[width=0.49\linewidth]{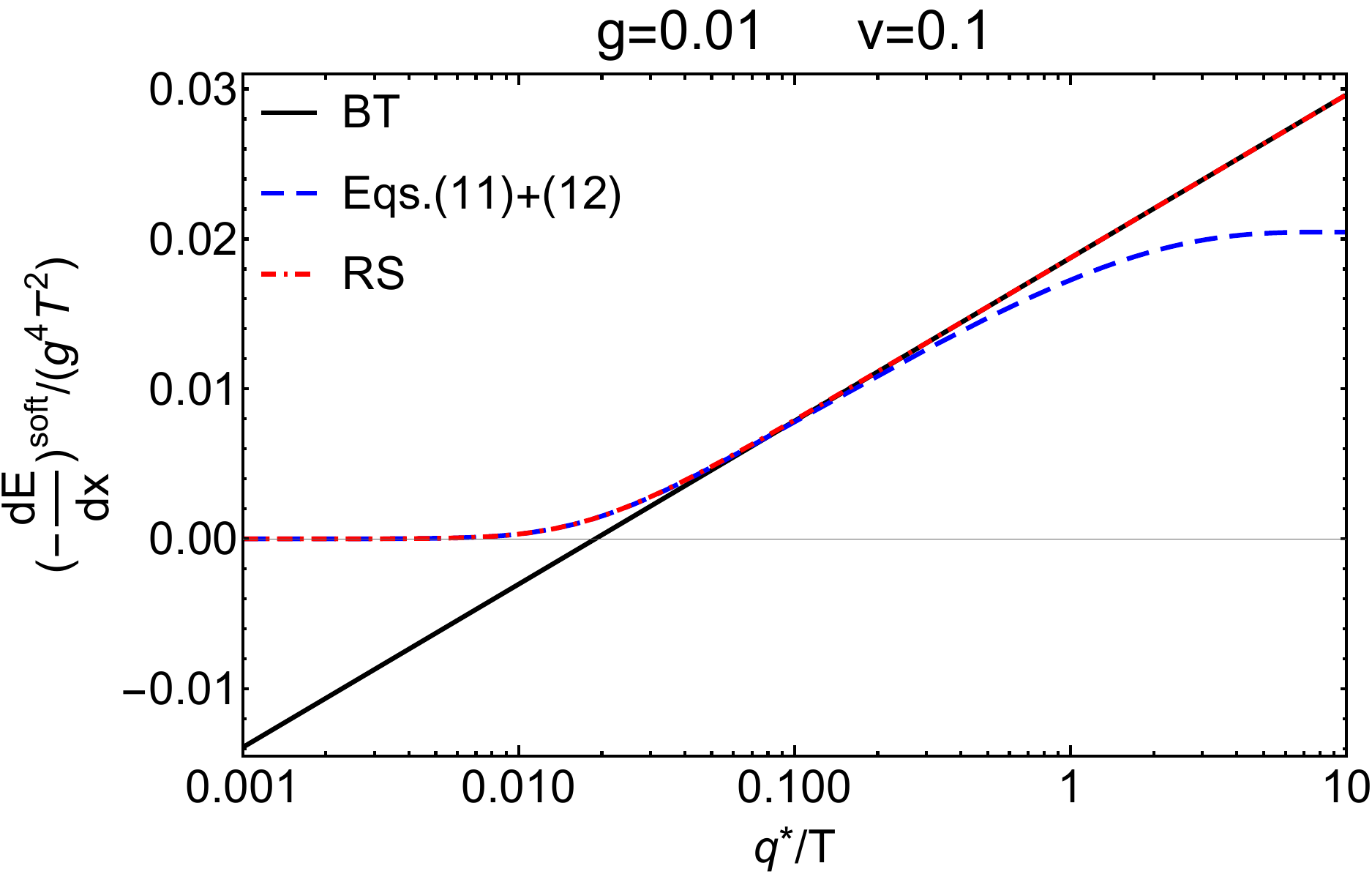}
\caption{Hard and soft contributions to the collisional energy loss as a function of $q^*/T$.
Results obtained from three different theoretical methods are presented for comparison.}
\label{comqs}
\end{center}
\end{figure}

Numerical results for the collisional energy loss from hard and soft processes are obtained based on different theoretical methods which can be found in Fig.~\ref{comqs}. In this figure, contributions from $s$ and $u$ channels\footnote{Their interference with $t$-channel scattering becomes negligible in our approximation $k, k^\prime \ll M_Q$.} have been included in hard scatterings between the heavy quark and gluons. These contributions can be easily obtained by using the bare quark propagator, and the result reads
\be\label{resu}
-\bigg(\frac{dE}{dx}\bigg)^{s+u}_{Qg} =\frac{4\pi\alpha_s^2  T^2}{3}\left(\frac{1}{v}-\frac{1-v^2}{2v^2}\ln \frac{1+v}{1-v}\right)\,.
\ee 
For comparison, we also present the corresponding results from Ref.~\cite{Braaten:1991we} [denoted as the Braaten-Thoma (BT) result] in Fig.~\ref{comqs}. In fact, the BT result can be obtained based on the RS result by further taking into account the weak-coupling limit $g\ll 1$, which guarantees $g T \ll q^* \ll T$ at high temperatures. For hard contribution, since $q^* \ll T$, one can set $q^* \rightarrow 0$ in Eq.~(\ref{changevah}) as long as this approximation does not lead to a divergence. For soft contribution, $ q^* \gg g T \sim m_D$ allows us to expand the RS result as given by Eqs. (\ref{elqqs}) and (\ref{elqgs}) and drop higher-order terms in the expansion. As a result, the remaining integrals in the RS result can be carried out analytically, which leads to the BT result. A key finding in the BT result is that the cutoff dependence is completely
canceled between the hard and soft contribution. On the contrary, when adding up the hard and soft contributions, the total energy loss based on the RS result still depends on the cutoff. One possible way to get rid of this ambiguity is to fix the cutoff by using the so-called minimal sensitivity; i.e., the energy loss is given by the minimum of $-d E/dx$ when varying $q^*$. 

As compared to the various results on $-d E/dx$ in QED obtained from the above-mentioned theoretical methods~\cite{Guo:2024mgh}, qualitatively similar behaviors are found in our complete QCD calculation. For hard scatterings, discrepancy at small $q^*/T$ originates from the different (bare or resummed) gluon propagators used in these different methods. A good agreement between our result and the RS result is found at large $q^*/T$, where the self-energy correction can be neglected. However, an unphysical energy loss appears in the BT result because the assumption $q^*\ll T$ does not hold anymore. For soft scatterings, our result agrees with the RS result very well provided that the momentum transfer is not large, i.e., $q^*\ll T$. As the cutoff becomes too small so that $g T\ll q^*$ cannot be satisfied, the BT result again gives a negative energy loss. All of these features are demonstrated in Fig.~\ref{comqs}.

According to the above discussions, in order to guarantee the validity of using a bare propagator, the cutoff cannot be too small for hard contributions. Therefore, disagreement among different results starts to show up when $q^*/T\sim g$. Furthermore, the cutoff cannot be too large for soft contributions, and discrepancy would emerge at $q^*/T\sim 1$, where the transferred momentum is no longer soft. After taking into account the further assumptions made in the BT result, we can expect that, for both hard and soft contributions, the different results become consistent in certain region of $q^*$ where the lower and upper bounds are determined by $q^*/T\gg g$, and $q^*/T\ll 1$, respectively. As we can see in Fig.~\ref{comqs}, in a region roughly given by $0.01<q^*/T<1$, numerical results obtained from different methods already show a reasonably good agreement. However, as the coupling constant increases, this region will shrink as a result of the increased lower bound. For large enough couplings, the lower bound can exceed the upper bound, and such a region eventually disappears. 


In addition, given the same parameter setup for the coupling constant and incident velocity\footnote{Our numerical results suggest that the boundary of the cutoff region is not sensitive to the specific values of the incident velocity.} as used in Ref.~\cite{Guo:2024mgh}, the cutoff region in which a consistent result can be obtained becomes narrower in QCD. This is mainly attributed to a reduced upper bound since no notable change is observed in the lower bound. We also point out that, for quark-quark scattering, this region is almost the same as that found in QED energy loss. Therefore, a quantitative difference in the upper bound must exist between the quark-quark and quark-gluon scatterings. This conclusion is also confirmed by our numerical results.

\begin{figure}[htbp]
\begin{center}
\includegraphics[width=0.49\linewidth]{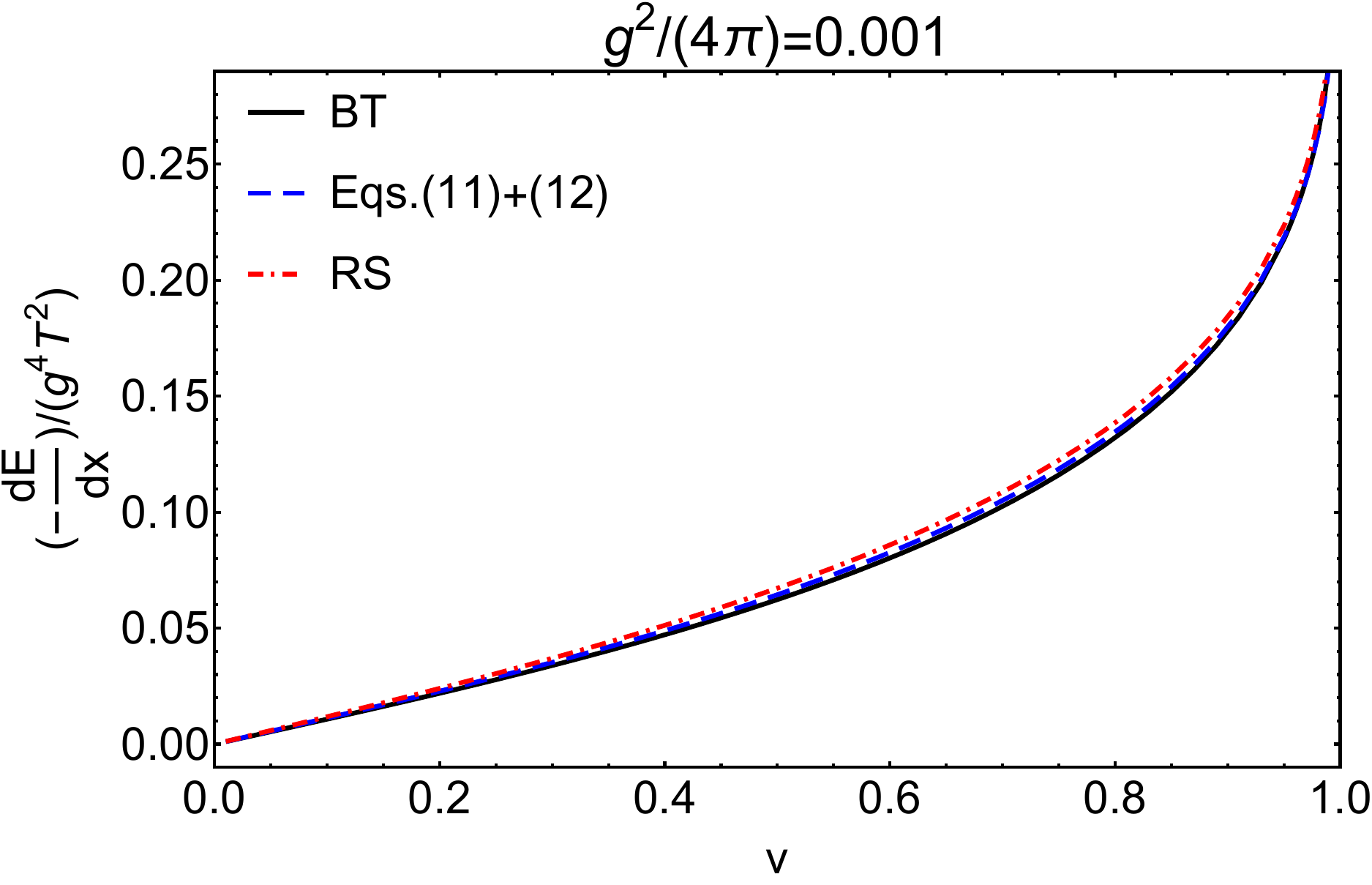}
\includegraphics[width=0.49\linewidth]{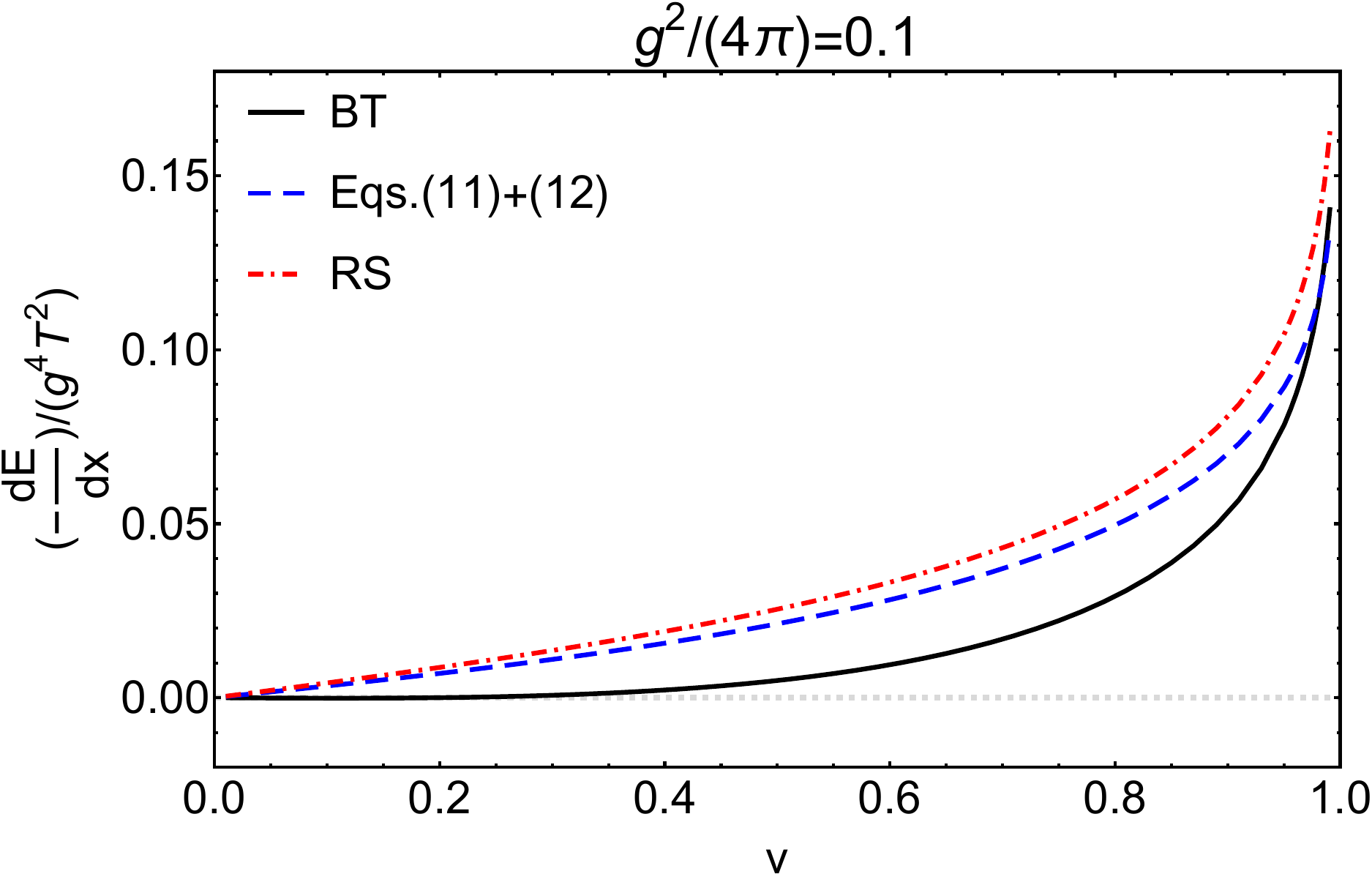}
\caption{Collisional energy loss of a heavy quark as a function of $v$ at different coupling constants. Results obtained from three theoretical methods are presented for comparison.}
\label{usdr2}
\end{center}
\end{figure}

Adding up the hard and soft contributions, the total energy losses of a heavy quark evaluated with different theoretical methods are shown in Fig.~\ref{usdr2}. By comparing the corresponding QED results from Ref.~\cite{Guo:2024mgh}, we can see that, although various results can converge at sufficiently small gauge coupling, such a convergence becomes much slower in QCD when decreasing the coupling constant. For example, given $e^2/(4\pi)=0.1$, the three different methods already give a very close result for the QED energy loss; however, a notable discrepancy is still presented in the QCD results according to Fig.~\ref{usdr2}. For a relatively large coupling constant, the BT result differs from the other two significantly, which also leads to a negative energy loss at small incident velocities. This behavior appears because of a term $\sim \ln(1/g)$ in the soft contribution. Our numerical evaluations suggest that, for a realistic value of the coupling constant in QCD, quantitative results of the collisional energy loss have a non-negligible dependence on the theoretical methods that one chooses. 

\section{Comments on the use of a resummed gluon propagator}\label{gauge}

Among the different theoretical methods to compute the collisional energy loss, the one with the use of a HTL resummed gluon propagator for arbitrary momentum transfer has a great advantage. In this method, the infrared divergence can be regulated without introducing an artificial cutoff, which will otherwise cause an unwanted  $q^*$ dependence in the collisional energy loss for any finite coupling constant. Furthermore, self-energy resummation can be smoothly included in the sense that a sharp transition from the hard to soft scatterings is absent. This feature is also important for studying the energy loss in a collisional plasma, where the collisions between medium partons play a role in the calculation of $-dE/dx$ through their modifications on the gluon self-energy. Although this method has already been adopted in some early works, the prescription with a resummed propagator cannot be trivially derived from the Feynman rules of thermal field theory. Therefore, some basic properties such as the gauge invariance cannot be automatically guaranteed. To extend our previous discussions on the gauge independence of the squared matrix element for quark-quark scattering, in this section, we will consider the related issues for quark-gluon scattering and further justify the use of such a prescription.

\subsection{On the gauge independence of the squared matrix element}\label{gain}

In this subsection, we will discuss the gauge independence of the squared matrix element $|{\cal{M}}|^{2}$ for the quark-gluon scattering processes. Although the gauge independence can be directly verified through a rather cumbersome calculation, we actually want to show the mechanism, especially for the case of using the resummed gluon propagator, how the gauge-dependent terms are canceled by each other, leaving $|{\cal{M}}|^{2}$ gauge independent. 

In general linear gauges, including covariant gauge, Coulomb
gauge, and temporal axial gauge, the gluon propagator $D^{\rho\sigma}(Q)$ can be written as
\be\label{generalpro}
-i D^{\rho\sigma}(Q)={\cal A} g^{\rho\sigma} + {\cal B} M^\rho M^\sigma +{\cal G} (M^\rho Q^\sigma +Q ^\rho M^\sigma) +{\cal G}^\prime Q^\rho Q^\sigma\, .
\ee
The first two terms are gauge independent which can be found in Eq.~(\ref{repro}). However, the specific forms of the last two terms associated with the coefficients ${\cal G}$ and ${\cal G}^\prime$ depend on the gauge we choose. We will show that these two coefficients do not show up in the final result of the squared matrix element for quark-gluon scattering; in other words, $| {\cal{M}}|^{2}$ is gauge independent regardless of which kind of propagator one uses for the exchanged gluon, the bare gluon propagator or the HTL resummed one, although the gauge independence of the former is already well known. 

Similar as the discussion on the quark-quark scattering in Ref.~\cite{Guo:2024mgh}, terms proportional to $Q^\rho$ in Eq.~(\ref{generalpro}) lead to a vanishing contribution to $i {\cal{M}}_{b}$ because of the following identity\footnote{According to Fig.~\ref{lod}, we use $i {\cal{M}}_{s}$ with $s=b,c,d$ to denote the matrix element for the $t$-, $s$-, and $u$-channel contribution in quark-gluon scattering, respectively. In addition, $i {\cal{M}}_{s} \equiv  i {\cal{M}}^{\mu \nu}_{s} \epsilon_{\mu}(K)\epsilon^*_{\nu} (K^\prime)$ and $i {\cal{M}} \equiv  i ({\cal{M}}_{b}+{\cal{M}}_{c}+{\cal{M}}_{d})$.}
\be\label{d0}
\bar{u}^{s'}(P') \displaystyle{\not} Q u^{s}(P)=\bar{u}^{s'}(P') (\displaystyle{\not}{P}-\displaystyle{\not}{P^\prime}) u^{s}(P)=0\, ,
\ee
which can be easily obtained based on the Dirac equations. Therefore, we can consider a simplified form of the gluon propagator
\be\label{generalpro2}
-i D^{\rho\sigma}(Q)={\cal A} g^{\rho\sigma} + {\cal B} M^\rho M^\sigma +{\cal G} M^\rho Q^\sigma \, .
\ee
Notice that the gauge-dependent term $\sim M^\rho Q^\sigma$ leads to the following Lorentz structure in $i {\cal{M}}_{b}$ when contracting with the three-gluon vertex ${\cal V}_{3g}^{\mu \nu \sigma}$:
\be\label{3gc}
Q_\sigma {\cal V}_{3g}^{\mu \nu \sigma} \equiv Q_\sigma[g^{\mu\sigma}(K-Q)^\nu+g^{\nu\sigma}(K^\prime+Q)^\mu+g^{\mu\nu}(-K-K^\prime)^\sigma]={K^\prime}^\mu {K^\prime}^\nu -K^\mu K^\nu\, .
\ee
For the initial- and final-state gluons, when we consider only the physical polarizations by replacing the polarization sums with $d^{\mu \nu}(K)$, the gauge-dependent part in the contraction $\left(i {\cal{M}}_{b}^{\mu\nu}\right) d_{\mu \mu^\prime}(K)d_{\nu \nu^\prime}(K^\prime)$ and its complex conjugate vanishes because of the transversality $K^\mu d_{\mu \mu^\prime}(K)={K^\prime}^\nu d_{\nu \nu^\prime}(K^\prime)=0$. This can be easily checked according to the definition
\be\label{d}
d_{\mu\nu} (K) = -g_{\mu\nu} + \frac{K_\mu N_\nu + K_\nu N_\mu}{K \cdot N} - N^2 \frac{K_\mu K_\nu}{(K \cdot N)^2}\, ,
\ee
with $N^\mu$ being an arbitrary four-vector. Thus, we have verified that the gauge-dependent terms in the gluon propagator do not contribute to the squared matrix element $|{\cal{M}}|^{2}$; therefore, it has no gauge dependence. Clearly, the above discussion works for both bare and resummed gluon propagators.

However, it turns to be less straightforward to show the gauge independence if the gluon polarization sums are replaced by $- g^{\mu \nu}(K)$. In this case, it is necessary to introduce the ghost field which, as we know in the use of a bare propagator, cancels unphysical gluon polarizations and ensures a gauge-independent $|{\cal{M}}|^{2}$. Given the fact that the resummed gluon propagator takes a much more complicated form as compared to the bare propagator, it is also worthwhile verifying an exact cancellation among those gauge-dependent contributions when the HTL resummed gluon propagator is used.\footnote{With the resummed gluon propagator, it remains to be seen whether or not the unphysical gluon polarizations are eliminated by the ghost field. 
We will discuss this issue in the next subsection.} 

We start by considering the interference terms\footnote{In this section, we consider a very general case without making any further assumption such as $k,k^\prime \ll M_Q$ like we used in the last section to compute $|{\cal M}|^2$. Therefore, we should also consider the interference terms.} between $i {\cal{M}}_{cd}\equiv i {\cal{M}}_{c}+i {\cal{M}}_{d}$ and $i {\cal{M}}_{b}$. For the gauge-dependent part proportional to ${\cal G}$ or ${\cal G}^*$, with the following contractions:
\be
(i {\cal{M}}_{cd})^{\mu \nu} K_{\mu} = - g^{2} f^{abc} t^{c}  \bar{u}^{s^{\prime}}(P^{\prime}) \gamma^{\nu} u^{s}(P)\,,\quad (i {\cal{M}}_{cd})^{\mu \nu} {K^\prime}_{\nu} = - g^{2} f^{abc} t^{c}  \bar{u}^{s^{\prime}}(P^{\prime}) \gamma^{\mu} u^{s}(P)\,,
\ee
we can get the same expression as given in Eq.~(\ref{d0}) thanks to the contraction in Eq.~(\ref{3gc}). Therefore, the gauge-dependent part in the interference terms vanishes. 

To proceed, we need only to consider the gauge-dependent terms in $| {\cal{M}}_b|^{2}$. According to the matrix element
\be
(i {\cal{M}}_{b})^{\mu \nu}= g^{2} f^{abc} t^{c}  \bar{u}^{s^{\prime}}(P^{\prime}) \gamma^{\rho} u^{s}(P)\big[({\cal A} g_{\rho \sigma} +{\cal B} M_\rho M_\sigma) {\cal V}_{3g}^{\mu \nu \sigma}+{\cal {G}} M_\rho({K^\prime}^{\mu}{K^\prime}^{\nu}-K^\mu K^\nu)\big]\,,
\ee
it can be shown that terms which are proportional to ${\cal A} {\cal G}^*$ or $ {\cal A}^*  {\cal G}$ in $| {\cal{M}}_b|^{2}$ also vanish due to the presence of the zero-valued $\bar{u}^{s'}(P')\displaystyle{\not} Q u^{s}(P)$. Notice that this expression arises as a consequence of the following contraction:
\be
({K^\prime}_{\mu}{K^\prime}_{\nu}-K_\mu K_\nu){\cal V}_{3g}^{\mu \nu \sigma}=K\cdot {K^\prime} ({K^\prime}^{\sigma}-K^\sigma)\, .
\ee
In fact, there are only two kinds of nonvanishing gauge-dependent contributions in $| {\cal{M}}_b|^{2}$. We denote them by $|{\cal{M}}_b|^{2}_{\rm g_1}$ and $| {\cal{M}}_b|^{2}_{\rm g_2}$. The former refers to a gauge-dependent term $\sim {\cal G} {\cal G}^*$, which reads
\be\label{g1}
| {\cal{M}}_b|^{2}_{\rm g_1}=-2 g^4 C_F N_c^2 {\rm Tr} \big[  \gamma^{0} (\displaystyle{\not}{P^\prime}+m) \gamma^{0} (\displaystyle{\not} P+m)\big] (K\cdot {K^\prime})^2 {\cal G}  {\cal G}^*\, .
\ee
It is familiar to us as it already appears when the bare gluon propagator is used. With a bare gluon propagator, it is well known that such a gauge-dependent term is canceled by the corresponding contribution in $| {\cal{M}}_{\rm ghost}|^{2}$. We emphasize that this cancellation is independent on the specific form of ${\cal G}$. Therefore, when using the resummed gluon propagator, the same cancellation occurs, although the coefficient ${\cal G}$ takes a more complicated form.

The other nonvanishing gauge-dependent contribution $\sim {\cal B} {\cal G}^*$ or $\sim {\cal B}^*  {\cal G}$ is new, because it shows up only when the resummed gluon propagator is used. This contribution is given by
\be\label{g2}
|{\cal{M}}_b|^{2}_{\rm g_2}=g^4 C_F N_c^2 {\rm Tr} \big[  \gamma^{0} (\displaystyle{\not}{P^\prime}+m) \gamma^{0} (\displaystyle{\not} P+m)\big] (K\cdot {K^\prime}) (k^\prime-k)  ( {\cal B} {\cal G}^*+ {\cal B}^*  {\cal G})\, .
\ee
After introducing the gluon-ghost interaction and using the HTL resummed propagator for the exchanged gluon, we can easily show that $|{\cal{M}}_b|^{2}_{\rm g_2}$ is exactly canceled by the term proportional to $ ( {\cal B} {\cal G}^*+ {\cal B}^*  {\cal G})$ in $|{\cal{M}}_{\rm ghost}|^{2}$ which comes out negative.\footnote{The gauge-dependent term $\sim {\cal A} {\cal G}^*$ or $\sim  {\cal A}^*  {\cal G}$ in $|{\cal{M}}_{\rm ghost}|^{2}$ also vanishes due to the identity Eq.~(\ref{d0}).}

Notice that the above discussions on the gauge independence do not rely on the specific forms of the scalar coefficients ${\cal A}$, ${\cal B}$, ${\cal G}$, and ${\cal G}^\prime$ as appear in Eq.~(\ref{generalpro}). Given this fact, when using the resummed propagator, the really new gauge-dependent structure that arises in $|{\cal M}|^2$  is given by Eq.~(\ref{g2}), because the term $\sim M^\rho M^\sigma$ in Eq.~(\ref{generalpro}) vanishes if we consider a bare propagator.

Based on the above discussions, we conclude that $|{\cal{M}}|^{2}$ is a gauge-independent quantity even if the resummed gluon propagator is used in the computation. Dropping the gauge-dependent terms in the resummed gluon propagator as we did in Eq.~(\ref{repro}) is, thus, verified to be valid. In addition, the gauge independence of $|{\cal{M}}|^{2}$ holds regardless of the way how one computes the squared matrix element; i.e., one can either consider only the physical polarizations by using $d^{\mu\nu}(K)$ to replace the polarization sums or introduce the ghost field when the simpler $-g^{\mu\nu}(K)$ is used. 

\subsection{The gluon polarization sums under the use of the resummed gluon propagator}

For the two different ways to compute the squared matrix element, we have shown the results are gauge independent. For convenience, we use $|{\cal{M}}|_{[d]}^{2}$ and $|{\cal{M}}|_{[g]}^{2}$ to denote the squared matrix element obtained with $d^{\mu\nu}(K)$ and $-g^{\mu\nu}(K)$ for gluon polarization sums, respectively. We know that, when using bare gluon propagator, the following identity holds
\be\label{eq}
|{\cal{M}}|_{[d]}^{2}=|{\cal{M}}|_{[g]}^{2}+|{\cal{M}}_{\rm ghost}|^{2}\, .
\ee
This identity implies that the ghost contribution can eliminate unphysical gluon polarizations, since the right-hand side of the above equation coincides with $|{\cal{M}}|_{[d]}^{2}$, which is obtained by using only the physical polarizations. Here, another issue we should further address is about the validity of the above equation when using the resummed propagator. In this subsection, we will show that Eq.~(\ref{eq}) is no longer true, in general, when the HTL resummed propagator is used. However, by self-consistently taking into account the HTL approximation, it still holds in an approximated way which makes our previous calculation of the squared matrix element meaningful.

We start by considering $|{\cal{M}}|_{[d]}^{2}$ which, according to Eq.~(\ref{d}), can be formally written as
\be
|{\cal{M}}|_{[d]}^{2}\equiv |{\cal{M}}|_{[g]}^{2}+\Delta |{\cal{M}}|_{[d]}^{2}\, ,
\ee
and the last term is given by
\ba\label{extra}
\Delta |{\cal{M}}|_{[d]}^{2} &=& (i\mathcal{M})^{\mu \nu} (i\mathcal{M})^{*\mu^{\prime} \nu^{\prime}} \left(-g_{\mu\mu^{\prime}}\frac{K_{\nu}^{\prime}N_{\nu^{\prime}}}{K^{\prime} \cdot N} -g_{\nu\nu^{\prime}}\frac{K_{\mu^{\prime}} N_{\mu}}{K \cdot N}
+\frac{ K_{\mu^{\prime}} K_{\nu}^{\prime} N_{\mu} N_{\nu^{\prime}}}{(K \cdot N)(K^{\prime} \cdot N)}
\right) \nonumber \\ 
&+& (i\mathcal{M})^{\mu \nu} (i\mathcal{M})^{*\mu^{\prime} \nu^{\prime}} K_{\nu^{\prime}}^{\prime}
\left( -g_{\mu\mu^{\prime}} + K_{\mu^{\prime}} \frac{N_{\mu}}{K \cdot N} \right) 
\left( \frac{N_{\nu}}{K^{\prime} \cdot N} - \frac{N^2 K_{\nu}^{\prime}}{(K^{\prime} \cdot N)^2} \right)\nonumber \\ 
&+ &(i\mathcal{M})^{\mu \nu} (i\mathcal{M})^{*\mu^{\prime} \nu^{\prime}} K_{\mu}
\left( -g_{\nu\nu^{\prime}} + K_{\nu}^{\prime} \frac{N_{\nu^{\prime}}}{K^{\prime} \cdot N} \right) 
\left( \frac{N_{\mu^{\prime}}}{K \cdot N} - \frac{N^2 K_{\mu^{\prime}}}{(K \cdot N)^2} \right) \nonumber \\
& +& (i\mathcal{M})^{\mu \nu} (i\mathcal{M})^{*\mu^{\prime} \nu^{\prime}} 
K_{\mu} K_{\nu^{\prime}}^{\prime} 
\left( \frac{N_{\mu^{\prime}}}{K \cdot N} - \frac{N^2 K_{\mu^{\prime}}}{(K \cdot N)^2} \right)
\left( \frac{N_{\nu}}{K^{\prime} \cdot N} - \frac{N^2 K_{\nu}^{\prime}}{(K^{\prime} \cdot N)^2} \right)\, .
\ea
The key point is to see whether or not $\Delta |{\cal{M}}|_{[d]}^{2}=|{\cal{M}}_{\rm ghost}|^{2}$. However, a direct numerical check shows that, in general, these two quantities differ from each other. We instead seek to find out some approximate equality under certain conditions.

We first look at the following contraction when using the resummed gluon propagator
\ba\label{con1}
 i\mathcal{M}^{\mu \nu} K_{\mu} 
&=& 
g^{2} f^{abc} t^{c} \bar{u}^{s^{\prime}}(P^{\prime}) \gamma^{\rho} u^{s}(P) 
\big[-{\cal A} {K^{\prime}}_{\rho} {K^{\prime}}^{\nu} - {\cal B} k^\prime M_{\rho} {K^{\prime}}^{\nu} -\big(1+{\cal A} Q^2\big) \delta^\nu_\rho
\nonumber \\
&+&{\cal B}\, M_{\rho} (\omega Q^\nu -Q^2 M^\nu)- {\cal G} (K^{\prime} \cdot Q) M_{\rho} {K^{\prime}}^{\nu}
\big]\, .
\ea
Unlike the use of a bare propagator, this expression is no longer proportional to ${K^{\prime}}^{\nu}$ due to several medium-induced contributions. Notice that, if we use the bare propagator, ${\cal B}=0$ and ${\cal A}=-1/Q^2$, only terms $\sim {K^{\prime}}^{\nu}$ survive. On the other hand, under the HTL approximation, those terms $\sim {K^{\prime}}^{\nu}$ dominate over the other terms in the above equation, and an approximated result reads
\be\label{con1app}
 i\mathcal{M}^{\mu \nu} K_{\mu} 
\approx
- g^{2}f^{abc} t^{c} \bar{u}^{s^{\prime}}(P^{\prime}) \gamma^{\rho} u^{s}(P) 
\big[{\cal A} {K^{\prime}}_{\rho} {K^{\prime}}^{\nu} + {\cal B} k^\prime M_{\rho} {K^{\prime}}^{\nu}
+ {\cal G} (K^{\prime} \cdot Q) M_{\rho} {K^{\prime}}^{\nu}
\big]\, .
\ee
In order to show the validity of the above approximation, recall that\footnote{If the energy of the incident quark is very large, $K^\prime$ could be higher order than $\sim T$ due to the extremely large momentum transfer. Then, one can simply use the bare gluon propagator, and Eq.~(\ref{eq}) holds.} the momenta $K$ and $K^\prime$ are at the order of $\sim T$, while the HTL gluon self-energy is $\sim g^2 T^2$. As the exchanged momentum $Q$ decreases from the hard scale $\sim T$ to the soft scale $\sim g T$, those terms in the square bracket in Eq.~(\ref{con1app}) will increase from ${\cal O}(1)$ to ${\cal O}(1/g^2)$. On the other hand, the magnitude of the dropped terms increases from ${\cal O}(g^2)$ to ${\cal O}(1)$. As a result, we can keep only terms $\sim {K^{\prime}}^{\nu}$ in Eq.~(\ref{con1}) in the weak-coupling limit, which is also consistent with the assumptions used in deriving the HTL resummed gluon propagator. 

Similarly, we can get the following contractions under the same approximation:  
\ba\label{con2app}
i \mathcal{M}^{* \mu^{\prime} \nu^{\prime}} K_{\nu^{\prime}}^{\prime} &\approx& -g^{2} f^{abd} t^{d} \bar{u}^{s}(P) \gamma^{\rho^{\prime}} u^{s^{\prime}}(P^{\prime})  \big[
{\cal A}^{*} K_{\rho^{\prime}}K^{\mu^{\prime}} + {\cal B}^{*} k M_{\rho^{\prime}} K^{\mu^{\prime}}
+ {\cal G}^{*} (K \cdot Q) M_{\rho^{\prime}} K^{\mu^{\prime}}
\big]\, ,  \nonumber\\
i \mathcal{M}^{* \mu^{\prime} \nu^{\prime}} K_{\mu^{\prime}} &\approx& -g^{2} f^{abd} t^{d} \bar{u}^{s}(P) \gamma^{\rho^{\prime}} u^{s^{\prime}}(P^{\prime})  \big[
{\cal A}^{*} K_{\rho^{\prime}}^{\prime} K^{\prime\nu^{\prime}} + {\cal B}^{*} k' M_{\rho^{\prime}} K^{\prime\nu^{\prime}}
+{\cal G}^{*} (K^{\prime} \cdot Q) M_{\rho^{\prime}} K^{\prime\nu^{\prime}}
\big]\, ,\nonumber \\
i \mathcal{M}^{\mu \nu} K_{\nu}^{\prime} &\approx&- g^{2} f^{abc} t^{c} \bar{u}^{s^{\prime}}(P^{\prime}) \gamma^{\rho} u^{s}(P)  \big[
{\cal A} K_{\rho}  K^{\mu} + {\cal B} k M_{\rho} K^{\mu}
+ {\cal G}( K \cdot Q) M_{\rho} K^{\mu}
\big]\, .
\ea

By repeatedly using the above contractions, many terms in Eq.~(\ref{extra}) vanish due to the on-shell condition $K^2=K{^\prime}^2=0$. Therefore, it can be simplified into 
\ba\label{extra2}
\Delta |{\cal{M}}|_{[d]}^{2} &=& (i\mathcal{M})^{\mu \nu} (i\mathcal{M})^{*\mu^{\prime} \nu^{\prime}} \left(-g_{\mu\mu^{\prime}}\frac{K_{\nu}^{\prime}N_{\nu^{\prime}}}{K^{\prime} \cdot N} -g_{\nu\nu^{\prime}}\frac{K_{\mu^{\prime}} N_{\mu}}{K \cdot N}
+\frac{ K_{\mu^{\prime}} K_{\nu}^{\prime} N_{\mu} N_{\nu^{\prime}}}{(K \cdot N)(K^{\prime} \cdot N)}
\right) \nonumber \\ 
&+& (i\mathcal{M})^{\mu \nu} (i\mathcal{M})^{*\mu^{\prime} \nu^{\prime}} 
\left(-g_{\nu\nu^{\prime}} \frac{ K_{\mu} N_{\mu^{\prime}}}{K \cdot N}  -g_{\mu\mu^{\prime}} \frac{K_{\nu^{\prime}}^{\prime} N_{\nu}}{K^{\prime} \cdot N}+\frac{ K_{\mu} K_{\nu^{\prime}}^{\prime} N_{\mu^{\prime}} N_{\nu}}{(K \cdot N)(K^{\prime} \cdot N)}\right)\, .
\ea
Furthermore, it is straightforward to show that, in each line of the above equation, contributions from the last two terms in the big brackets are canceled by each other, and the nonvanishing contribution from the second line is just the complex conjugate of the first line. Consequently, the final expression of $\Delta |{\cal{M}}|_{[d]}^{2}$ reads
\ba\label{simple}
\Delta |{\cal{M}}|_{[d]}^{2} &\approx& -g^4 C_F N_c^2 {\rm Tr} \big[  \gamma^{\rho^{\prime}} (\displaystyle{\not}{P^\prime} +m)\gamma^{\rho} (\displaystyle{\not} P+m) \big]\big[
{\cal A} K_{\rho}  + {\cal B} k M_{\rho} 
+ {\cal G}( K \cdot Q) M_{\rho} 
\big] \nonumber \\ 
&\times &\big[
{\cal A}^{*} K_{\rho^{\prime}}^{\prime}  + {\cal B}^{*} k' M_{\rho^{\prime}} 
+{\cal G}^{*} (K^{\prime} \cdot Q) M_{\rho^{\prime}} 
\big]+(K \leftrightarrow -K^\prime)\, ,
\ea
which is exactly the same as $|{\cal{M}}_{\rm ghost}|^{2}$. Therefore, contributions from unphysically polarized gluons are canceled by the ghost field as expected. At this point, we can conclude that, in the HTL approximation, Eq.~(\ref{eq}) holds when the resummed gluon propagator is used in computing $|{\cal{M}}|^{2}$ for the QCD Compton scattering. 

Finally, we point out that, when using the resummed gluon propagator and abandoning the approximations given in Eqs.~(\ref{con1app}) and (\ref{con2app}), there is an ambiguity in the determination of $|{\cal{M}}|_{[d]}^{2}$, because the result depends on the four-vector $N^\mu$ as introduced in Eq.~(\ref{d}). In addition, the difference between $\Delta |{\cal{M}}|_{[d]}^{2}$ and $|{\cal{M}}_{\rm ghost}|^{2}$ is found to be $\sim C_F N_c^2$, which also depends on the specific form of $N^\mu$, while Eq.~(\ref{eq}) can be always satisfied by contributions $\sim C_F$ .

\section{The collisional energy loss of a heavy quark in the QGP with a BGK collisional kernel}\label{elwithcolli}

The dissipation of energy can occur in a collisionless plasma due to Landau damping which leads to an imaginary part in the gluon self-energy in Eq.~(\ref{pi}). However, this process does not produce entropy. As another important source for dissipation, collisions between the constituent partons of the hot QGP are responsible for driving the system to the equilibrium state of maximum entropy. With these collisions, collisional damping of the quasiparticle modes emerges in addition to a modification on the Landau damping. Accordingly, collective behaviors exhibit some new features in a collisional plasma~\cite{Carrington:2003je,Schenke:2006xu,Zhao:2023mrz}. It is also important to study the collision effect on the relevant physical quantities, such as the energy loss of a heavy quark. We also note that collisions become negligible in the small-coupling limit where the kinetic energy of the partons is much larger than the potential energy of two neighboring partons. On the other hand, at moderate coupling constants, collision effect may play an important role on the collisional energy loss of a heavy quark. Given our method to compute the energy loss, a natural way to incorporate the collision effect is to consider the corresponding modifications on the resummed gluon propagator. To do so, we need to derive the gluon self-energy by taking into account the collisions between medium partons, which can be done by using the kinetic theory with a specified collisional kernel.

The theoretical method to compute the collisional energy loss as adopted in this work turns out to be more suitable for being used in a collisional plasma. Unlike the other two methods~\cite{Braaten:1991we,Romatschke:2004au}, where the collision effect never appears in hard scatterings due to the use of a bare propagator, the influence of collisions between thermal partons will enter into all scattering processes in our approach, regardless of the magnitudes of the transferred momenta. Although, for hard processes, self-energy insertion into the bare gluon propagator is not expected to induce a significant modification on the energy loss, treating both the hard and soft scatterings in a unified framework will enable us to self-consistently include the effect of collisions, which is also free of any artificial cutoff.

As an extension of our previous work~\cite{Guo:2024mgh} to full QCD, we employ the same BGK collisional kernel as given by
\be\label{bgk}
{\cal C}({\bf k},X)=-\nu \Big[f({\bf k},X)-\frac{\int_{\bf k}f({\bf k},X)}{\int_{\bf k}n_{B/F}(k)}n_{B/F}(k)\Big]\, .
\ee
In the above equation, $\int_{\bf k}\equiv \int d^3{\bf k}/(2\pi)^3$ and $f({\bf k}, X)=f ({\bf{k}})+\delta f({\bf k}, X)$. The fluctuation $\delta f({\bf k}, X)$ presents a slight deviation of the distribution function from its homogeneous values $f(\bf{k})$. In addition, $\nu$ denotes the collision rate, which is very crucial for quantitative investigations on the collision effects. However, determining the collision rate appearing in the BGK collisional kernel turns to be very challenging because Eq.~(\ref{bgk}) serves as only a phenomenological model for equilibration which cannot be derived from first principles. On the other hand, calculation of the collision rate in a perturbative framework finds $\nu \approx 5.2 \alpha_s^2 T \log (0.25 \alpha_s^{-1})$~\cite{Thoma:1993vs}. However, this result leads to a negative collision rate for realistic values of the coupling constant, $\alpha_s\approx0.2- 0.4$, that relevant for heavy-ion collisions. To overcome this problem, following Refs.~\cite{Schenke:2006xu,Kapusta:1991qp}, we insert a constant $c$ into the logarithm; as a result, the dimensionless collision rate reads
\be\label{nu}
{\tilde \nu}\equiv \nu/m_D \approx 1.27 \alpha_s^{3/2} \ln (c+ 0.25/\alpha_s)\, .
\ee
In the numerical evaluations, the value of $c$ varies from $1$ to $2$ to account for the uncertainties of the above parametrization. As we can see, the parameter $c$ can be neglected in the small-coupling limit $\alpha_s \rightarrow 0$, where the result in Ref.~\cite{Thoma:1993vs} is expected to work. For a typical value of the coupling constant $\alpha_s=0.3$, the collision rate $\nu/m_D \approx 0.13 - 0.22$ which is also consistent with the values commonly used. Nevertheless, we have to admit that such a parametrization in Eq.~(\ref{nu}) cannot be strictly justified at present. In order to get meaningful results of the collision rate even for moderate coupling constants, one could revisit the calculation in Ref.~\cite{Thoma:1993vs} by improving upon the techniques used where necessary. This is certainly a nontrivial work and beyond the scope of this paper. 

Given the collisional kernel in Eq.~(\ref{bgk}), we can solve the linearized kinetic equation for the fluctuation $\delta f({\bf k}, X)$ which determines the induced current $J_{{\rm ind}}^\mu(Q)$. Then the gluon self-energy $\Pi^{\mu\nu} (Q)$ can be derived by functional differentiation of $J_{{\rm ind}}^\mu(Q)$ with respect to the gauge field. More details about this derivation can be found in Refs.~\cite{Carrington:2003je,Schenke:2006xu,Zhao:2023mrz}. We point out only that, when considering the thermal equilibrium distributions, i.e., $f({\bf{k}})=2N_f n_F(k)+6 n_B(k)$, the resulting gluon self-energy retains the same Lorentz structure as its counterpart in the collisionless limit. Therefore, it becomes quite straightforward to get the resummed gluon propagator via the Dyson-Schwinger equation. When compared to Eq.~(\ref{repro}), the modification due to the collisions on the resummed gluon propagator amounts to only the following changes on the transverse and longitudinal self-energies:
\ba\label{pitl}
 \Pi_T ({\hat {\omega}},{\hat {\nu}})&=& \frac{m_D^2}{4} {\hat{\omega}} \bigg[2z+(z^2-1)\ln\frac{z - 1}{z + 1}\bigg]\, ,\nonumber\\
 \Pi_L ({\hat {\omega}},{\hat {\nu}}) &=&-\frac{m_D^2}{2} \frac{1 }{ \mathcal{W}(\hat \omega,\hat\nu )}\bigg(2+z {\ln\frac{z-1}{z+1}}\bigg) \,,
\ea
where
\be
\mathcal{W}(\hat \omega,\hat\nu )=1-\frac{i \uh}{2}\int_{-1}^{1}\mathrm{d} x\frac{1}{\oh-x+ i\uh}=1+\frac{i\uh}{2} \ln\frac{z-1}{z+1}\, ,
\ee
with $z\equiv {\hat {\omega}} +i {\hat {\nu}}$ and ${\hat {\nu}}=\nu/q$. It is obvious that our previous discussions on the gauge independence as well as the gluon polarization sums still hold in the presence of the BGK collisional kernel. As a result, one can directly use Eqs.~(\ref{elqq}) and (\ref{elqg}) to evaluate the energy loss and investigate the influence of collisions between medium partons on $- dE/d x$. 

\begin{figure}[htbp]
\begin{center}
\includegraphics[width=0.49\linewidth]{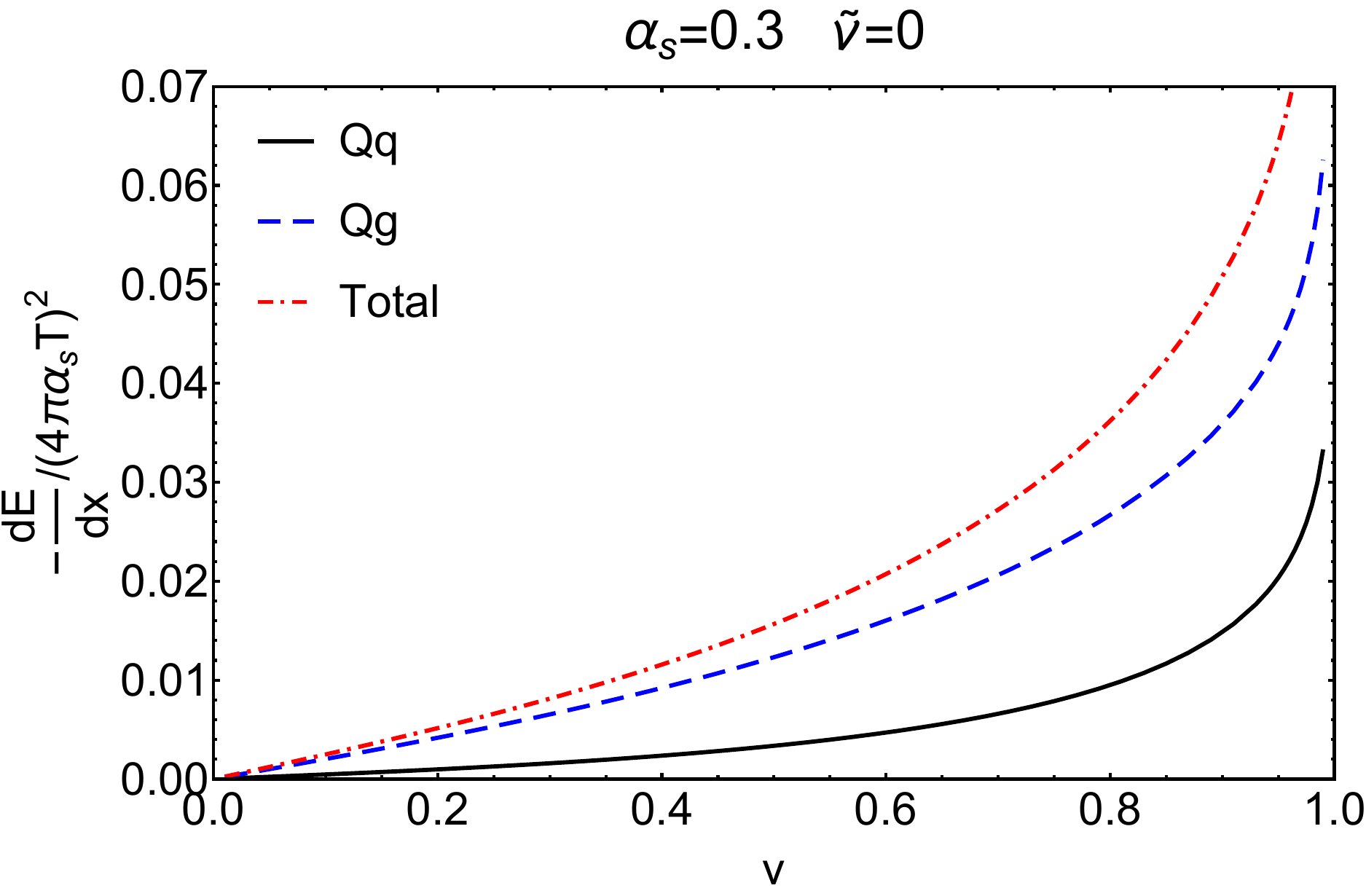}
\includegraphics[width=0.49\linewidth]{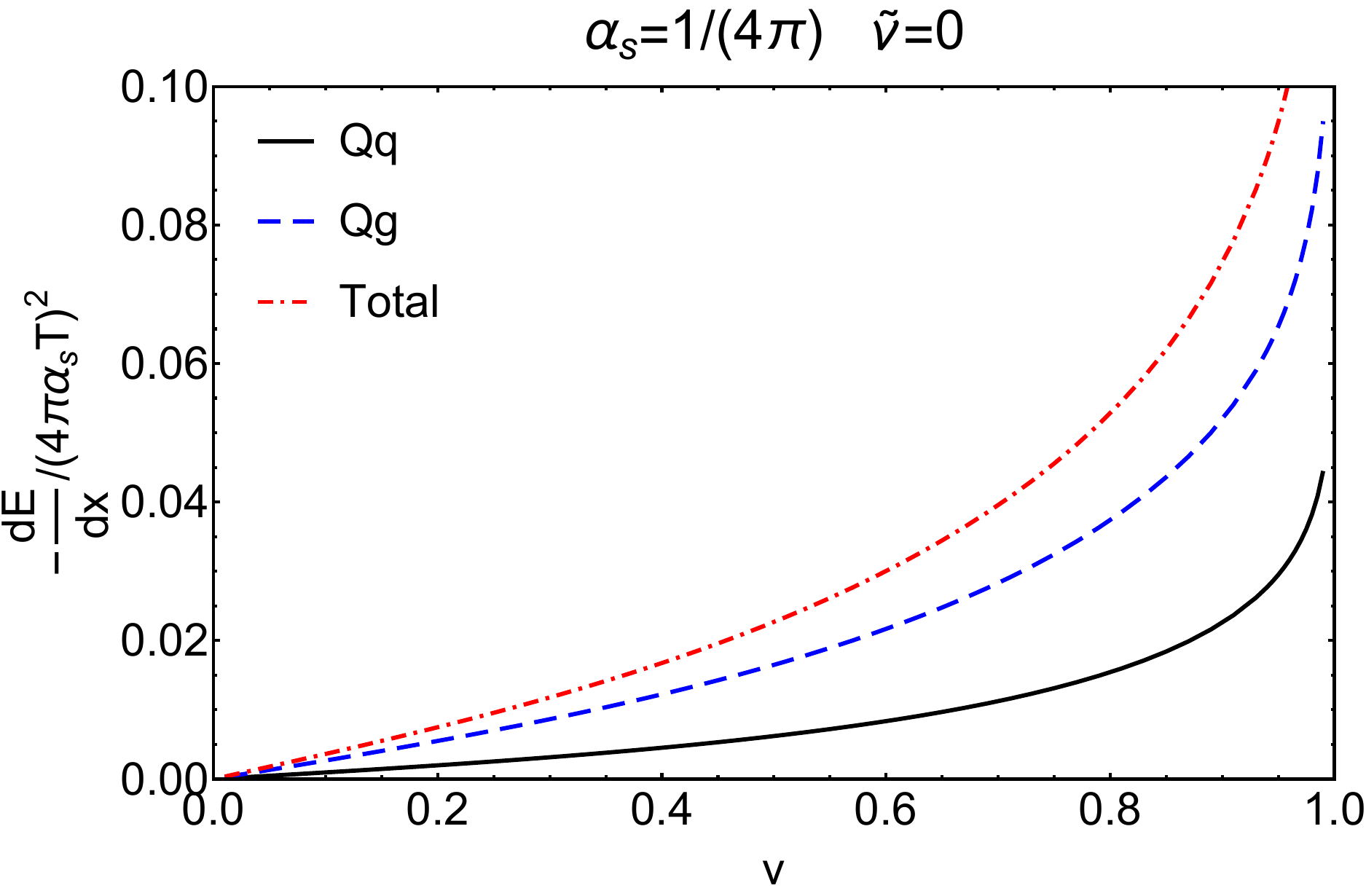}
\caption{The heavy-quark energy loss as a function of the incident velocity at different coupling constants. Besides the total energy loss, contributions from quark-quark and quark-gluon scatterings are also presented separately.}
\label{usdr}
\end{center}
\end{figure}

The numerical results of the heavy-quark energy loss at different coupling constants are presented in Fig.~\ref{usdr}, where the results have been scaled by a factor of $(4\pi \alpha_s T)^{-2}$ so that there is no explicit $T$ dependence in the dimensionless energy loss. In general, the energy loss strongly depends on the incident velocity $v$, and a dramatic increase of $-d E/d x$ is found at very large velocities. When comparing with the results in Ref.~\cite{Guo:2024mgh}, where only the contribution from quark-quark scattering is considered, a much larger portion of the total energy loss actually comes from the quark-gluon scattering. Beside a nonvanishing $s$- and $u$-channel contribution in QCD, the $t$-channel quark-gluon scattering also significantly contributes to the collisional energy loss. According to Eq.~(\ref{resu}), the former leads to a constant shift of the dimensionless energy loss at any given incident velocity $v$, while the result of $t$-channel scattering has an extra dependence on the coupling constant. Therefore, the ratio of $-d E/d x$ from quark-quark scattering to that from quark-gluon scattering exhibits a complicated dependence on both $\alpha_s$ and $v$. We have checked that, in the weak-coupling limit, the value we obtained for this ratio is $\sim 0.38$, agrees with BT result, and exhibits a very weak $v$ dependence. For typical values of the coupling constant near the critical temperature $\alpha_s=0.3$, this ratio gets a notable increase with $v$. However, even at very large incident velocity, the contribution from quark-gluon scattering is still dominant which is about twice the contribution from quark-quark scattering. This finding indicates the necessity of taking into account the quark-gluon scattering in a rigorous way as we did in this work.

\begin{figure}[htbp]
\begin{center}
\includegraphics[width=0.49\linewidth]{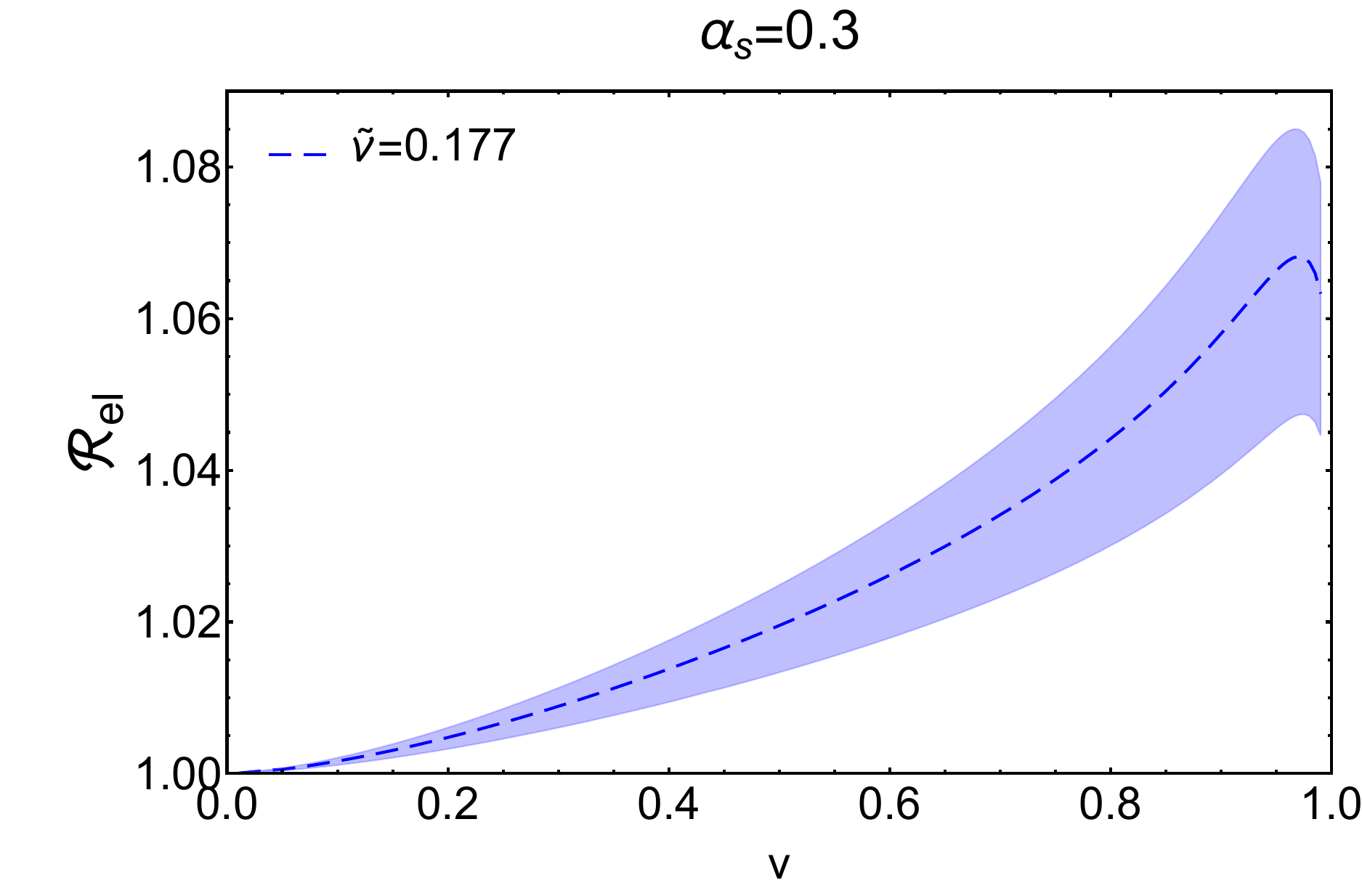}
\includegraphics[width=0.49\linewidth]{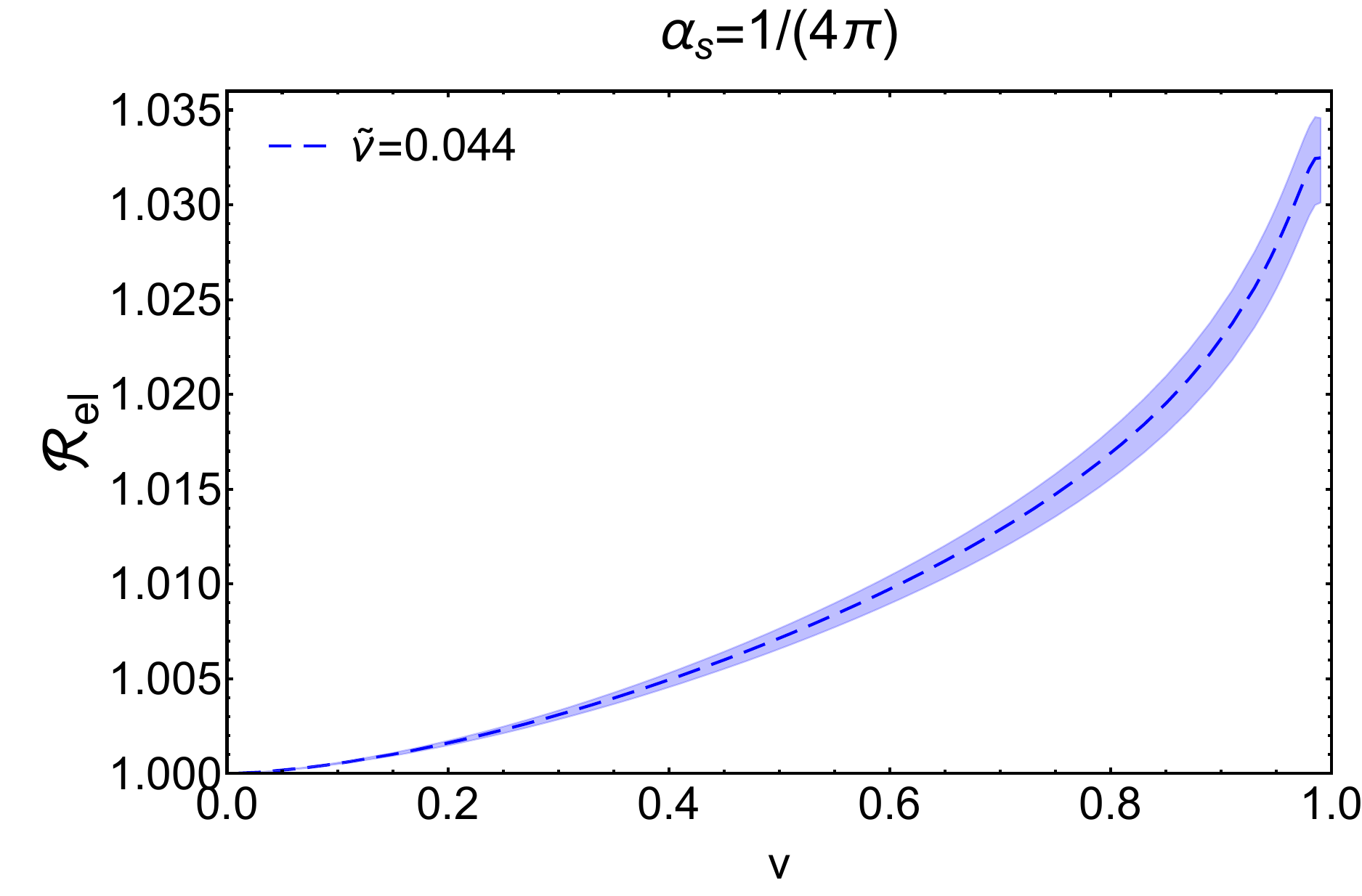}
\caption{The energy loss ratio as a function of the heavy-quark velocity at different coupling constants or collision rates. The bands correspond to varying the parameter $c$ in Eq.~(\ref{nu}) from $1$ to $2$.}
\label{ravsv}
\end{center}
\end{figure}

Another focus in this work is the collision-induced correction to the heavy-quark energy loss. A clearer way to illustrate this correction is to look at the ratio of $-dE/dx$ with and without the collisions. We use ${\cal R}_{\rm el}$ to denote this ratio, which can be expressed as
\be
{\cal R}_{\rm el} \equiv \bigg(\frac{d E}{d x}\bigg)_{\nu\neq 0}\bigg/\bigg(\frac{d E}{d x}\bigg)_{\nu= 0}\,.
\ee
Notice that, in Ref.~\cite{Guo:2024mgh}, such an energy loss ratio in QCD was estimated based on a QED calculation of the collisional energy loss. This estimation strongly relies on an assumption of an equal energy loss ratio for quark-quark and quark-gluon scatterings. With our complete QCD calculation, more reliable results can be provided, which are presented in Fig.~\ref{ravsv}. In our numerical evaluations, we choose $c=1.5$ to determine the collision rate. To account for the uncertainties of the parametrization in Eq.~(\ref{nu}), we also vary the value of $c$ from 1 to 2, which corresponds to a band in the plots. As we can see, the energy loss ratio is sensitive to the values of the coupling constant. Taking $\alpha_s=1/(4\pi)$, the collision rate in Eq.~(\ref{nu}) is given by ${\tilde \nu}\approx 0.044$, determined at $c=1.5$. For such a small $\nu$, collisions between the medium partons do not play a notable role on the heavy-quark energy loss, because the collision-induced correction to $-d E/d x$ does not exceed $\sim 4 \%$, which also agrees very well with the estimation in Ref.~\cite{Guo:2024mgh}. However, for a coupling constant $\alpha_s =0.3$, corresponding to ${\tilde \nu} \approx 0.177$, the corrections can reach $\sim 8 \%$, compared to $\sim 11 \%$ as found in Ref.~\cite{Guo:2024mgh}. Despite a slight decrease in magnitude, our new result also indicates a moderate collision effect which enhances the collisional energy loss, especially at large incident velocities. For quantitative accuracy, taking into account the collision effect is still important when evaluating the heavy-quark energy loss near the critical temperature. 

The above discussions suggest that the energy loss ratio obtained from the complete QCD calculation is reduced as compared to the corresponding estimations in Ref.~\cite{Guo:2024mgh}, although the differences are not significant even at a moderate coupling constant. On the one hand, the assumption of an equal energy loss ratio for quark-quark and quark-gluon scatterings used in Ref.~\cite{Guo:2024mgh} is, thus, justified to some extent. On the other hand, it also implies that the collision effect is less accentuated after including the quark-gluon scattering. This conclusion requires that the energy loss ratio for quark-gluon scattering must be smaller than that for quark-quark scattering. In fact, we have numerically checked that, when considering only the $t$-channel contributions, the collision-induced correction is even larger for quark-gluon scattering. Therefore, the reduced collision effect is entirely attributed to the contributions from the $s$ and $u$ channels as they are unaffected by the collisions between the medium partons. 

The results presented in Fig.~\ref{ravsv} qualitatively agree with those obtained in previous works~\cite{Han:2017nfz,YousufJamal:2019pen}. However, given a similar collision rate, our energy loss ratio is significantly lower than the previously found. The overestimated collision effect in the previous works stems from the neglect of the hard contributions which, as compared to the soft ones, are less affected by the collisions between medium partons. Notice that the theoretical formula adopted in Refs.~\cite{Han:2017nfz,YousufJamal:2019pen} is applicable only to computing soft scatterings; therefore, the resulting $-dE/dx$ depends on a cutoff which regulates ultraviolet divergence by setting a maximum for transferred momenta. On the contrary, as an important advantage of the theoretical method in this work, our results are free of any artificial cutoff.

The increased energy loss in a collisional plasma indicates a rise of the scattering probability between incident heavy-quark and medium partons. However, it is not obvious to see the underlying mechanism for such an increase, because a nonzero collision rate $\nu$ gives rise to a highly nontrivial modification on the transverse and longitudinal gluon propagators. An intuitive way to understand the above results is to look at the soft contributions in the energy loss. According to the discussion under Eq.~(\ref{changeva4}), the integration of distribution function over $k$ is not affected by the collisions between medium partons, and the collision effect appears only in the remaining integrals over $\omega$ and $q$ as given by Eqs.~(\ref{elqqs}) and (\ref{elqgs}). Considering a collisionless QGP in the weak-coupling limit, integrating over $\omega$ and $q$ just leads to the dimensionless energy loss $\sim s (v) \log (q^*/m_D)$, where $s(v)$ is a function of the heavy-quark velocity~\cite{Braaten:1991we}. Naively, we can expect that, to get an enhancement in the energy loss, the screening mass in the logarithm would be effectively decreased by a nonzero collision rate $\nu$. In addition, the decrease of screening mass also depends on the incident velocity $v$, since the energy loss ratio shows a rapid change when varying $v$. To fully understand the complicated dependence of $-d E/ dx$ on the collision rate and incident velocity, one should exactly carry out the integrals over $\omega$ and $q$ in a collisional QGP; however, this turns out to be very difficult even in the weak-coupling limit.

\begin{figure}[htbp]
\begin{center}
\includegraphics[width=0.49\linewidth]{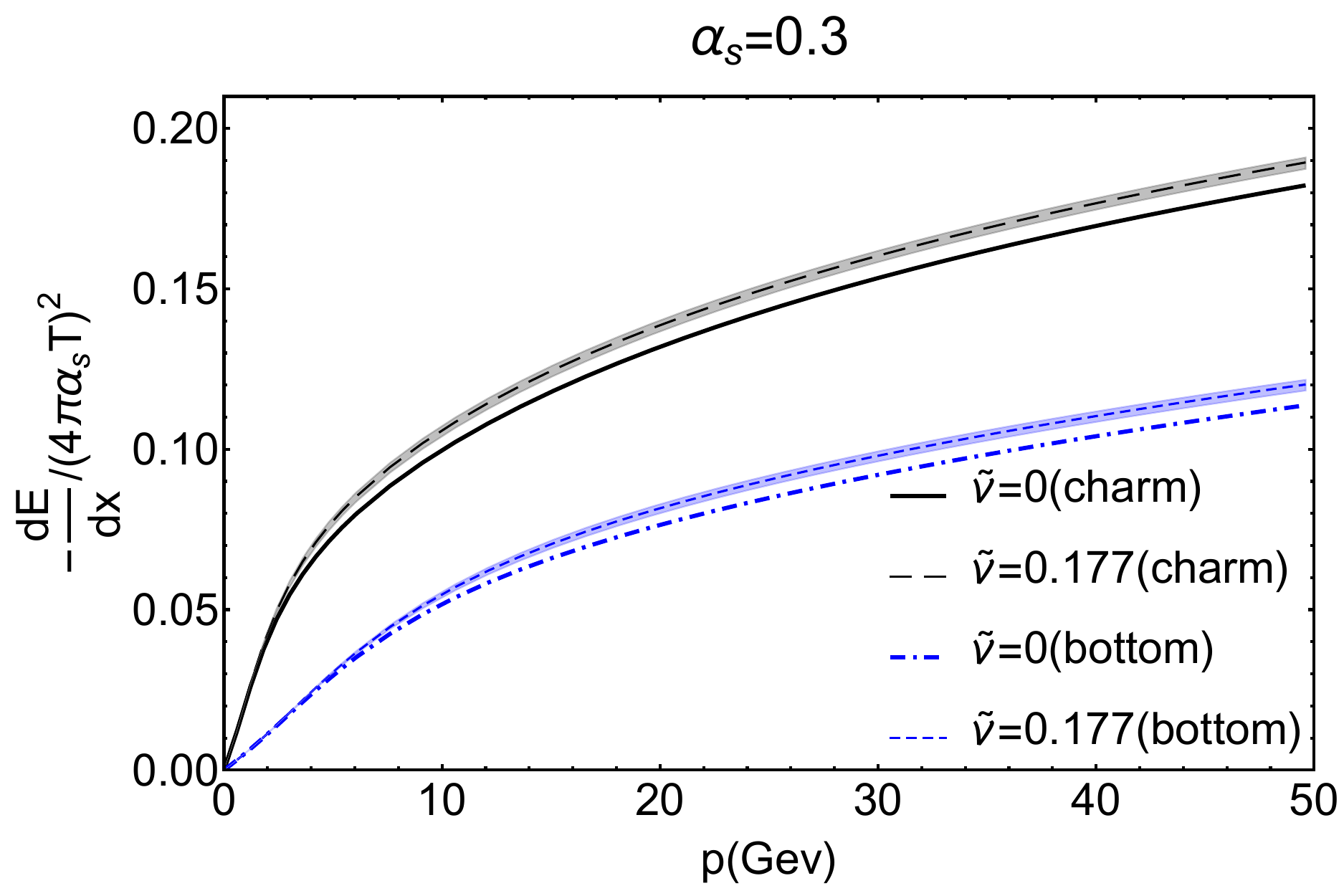}
\includegraphics[width=0.49\linewidth]{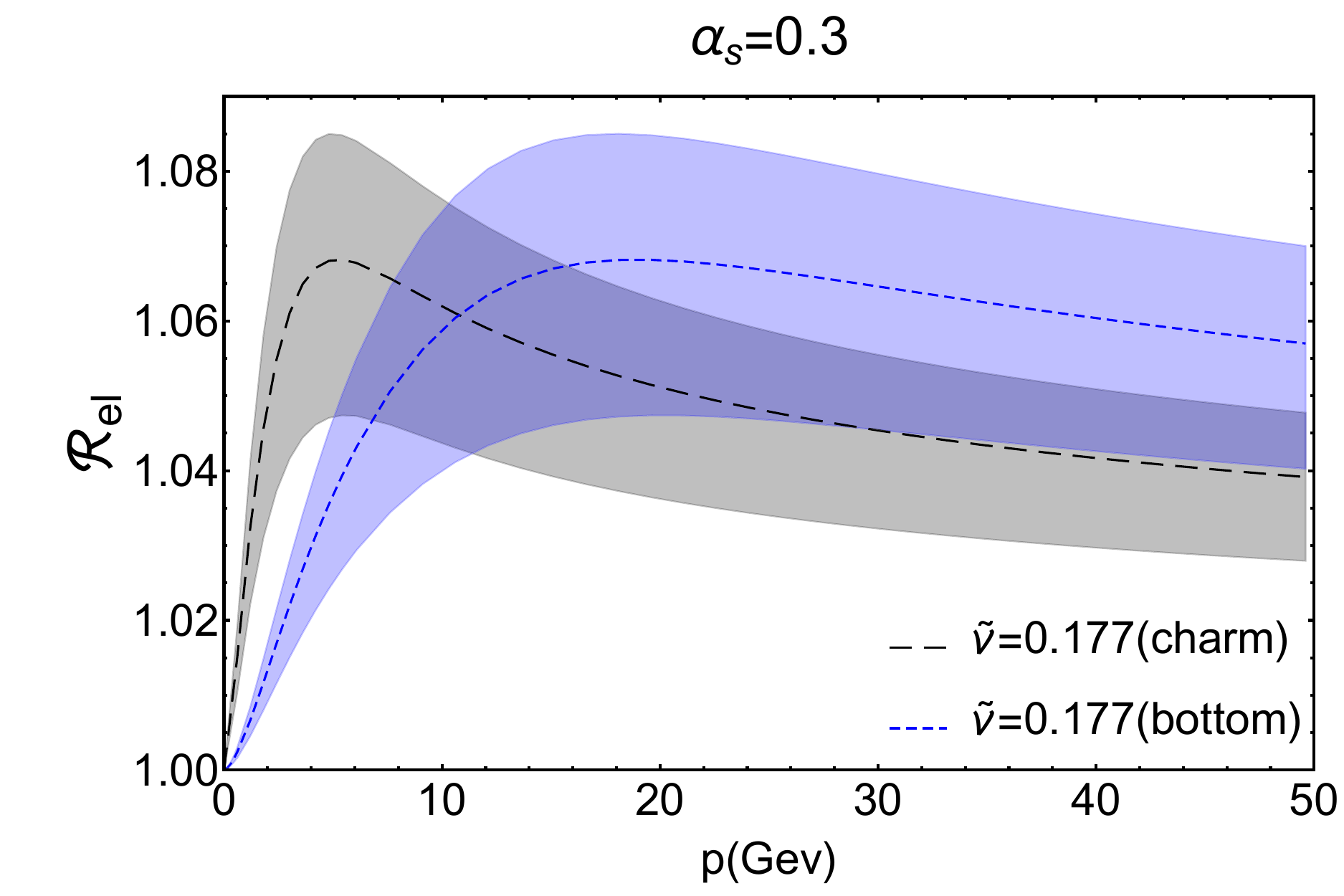}
\caption{Left: comparisons of the momentum dependence of the energy loss with and without collisions for charm and bottom quarks at $\alpha_s=0.3$.  Right: the energy loss ratio as a function of the momentum for charm and bottom quarks at $\alpha_s=0.3$. The bands correspond to varying the parameter $c$ in Eq.~(\ref{nu}) from $1$ to $2$.}
\label{elcandb}
\end{center}
\end{figure}

In Fig.~\ref{elcandb}, we plot the dimensionless energy loss as well as the energy loss ratio as a function of the heavy-quark momentum $p$. We consider a realistic value of the coupling constant $\alpha_s=0.3$ and choose $M_c =1.3\,{\rm GeV}$ and $M_b =4.7\,{\rm GeV}$ for the charm and bottom quark mass, respectively. The qualitative behaviors of the $p$ dependence are found to be similar as our previous results in Ref.~\cite{Guo:2024mgh}. Notice that the energy loss increases monotonically with the incident velocity; for a given $p=v M_Q/\sqrt{1-v^2}$, the bottom quark with a smaller velocity loses less energy than the charm quark. This is observed in our results where the complete QCD calculation leads to a dramatic enhancement of energy loss as compared to Ref.~\cite{Guo:2024mgh}, although the ratio between the bottom quark energy loss and the charm quark energy loss does not change significantly. In addition, the collision-induced corrections are more pronounced in the large momentum region for the bottom quark energy loss, while in the small momentum region, the collision effect plays a more important role on the charm quark energy loss. The maximum of the energy loss ratio is found to be consistent with the result presented in Fig.~\ref{ravsv}. 

Finally, we point out that including the strange quarks does not change the above conclusions with $N_f=2$. Numerically, we find that the energy loss ratio is almost unaffected. Although contributions to the energy loss due to quark-quark scattering increase with one more thermally active flavor, quark-gluon scattering is still dominant in the heavy-quark energy loss. We also need to emphasize that, in our evaluations, the values of the collision rate are provided based on a phenomenological parametrization. A quantitatively accurate determination of the collision rate must be carried out in the future, which is crucial to have a conclusive assessment on the heavy-quark energy loss in a collisional QCD plasma.

\section{Conclusions and Outlook}\label{con}

In this work, we presented a complete QCD calculation on the heavy-quark energy loss in a collisional QGP. By employing the HTL resummed gluon propagator, we obtained the interaction rate for the elastic scatterings between the incident heavy-quark and the medium partons. The collision effect was encoded in the resummed propagator via the collisionally modified gluon self-energy which was obtained by solving the linearized kinetic equation in the presence of the BGK collisional kernel.

We first justified the theoretical method to compute the collisional energy loss. It was essentially about the validity of using the resummed gluon propagator for arbitrary momentum transfer which treated the hard and soft processes in a unified framework. On the one hand, we demonstrated 
by a direct comparison that our result coincided with the RS result in the limit of either hard or soft momentum transfers. When integrating over all the possible transferred momenta, different methods led to an identical energy loss in the small-coupling limit; however, for finite coupling constants, notable discrepancy appeared. This discrepancy became more relevant in QCD due to a moderate coupling constant.  
On the other hand, by including the quark-gluon scattering, we explicitly showed the gauge independence of the squared matrix element for full QCD and demonstrated the elimination of unphysical gluon polarizations by the ghost contribution under the HTL approximation. These conclusions further established a solid foundation for using the resummed gluon propagator to compute the interaction rate. 

Based on the above theoretical method, we studied the collisional energy loss of a heavy quark traversing through the QGP with a focus on how the collisions between the medium partons could affect the results. As an extension of our previous work on the QED energy loss, the complete QCD calculation showed that a very large portion of the collisional energy loss actually came from the quark-gluon scattering and, therefore,  
could provide a quantitatively reliable result for the heavy-quark energy loss although the uncertainty originated from the parametrization of the collision rate in BGK collisional kernel needs to be further studied. As for the influence of the collision effect, the heavy-quark energy loss was increased in a collisional QGP, and such an increase became non-negligible for large incident velocities and moderate coupling constants. As compared to the estimate in Ref.~\cite{Guo:2024mgh}, the collision effect as observed in this work was less accentuated. The maximum of the collision-induced correction to energy loss was reduced from $\sim 11 \%$ to $\sim 8 \%$ at the gauge coupling $\alpha_s=0.3$, which indicated that the quark-gluon scattering was less affected by the collisions when comparing with the quark-quark scattering. In addition, the collision effect also exhibited a flavor dependence as discovered before. At large momenta, the corrections increased with increasing heavy-quark mass while at small momenta, an opposite trend was observed.

Finally, the theoretical method of computing the collisional energy loss as used in this work can be adopted in many other studies where physical processes involving soft momentum exchanges are relevant. More importantly, instead of considering the collision effect on the physical quantities in consideration, other medium effects can be also taken into account in a similar manner, namely, by using an effective propagator that resums an infinite series of the self-energy contribution modified by various medium effects. Another related work is to study the jet quenching parameter by introducing a background field which describes a nontrivial temperature dependence of the order parameter for the deconfining phase transition~\cite{Dumitru:2010mj,Dumitru:2012fw,Guo:2014zra}. A nonvanishing background field corresponds to the partial deconfinement of the QGP, which gives rise to modifications on the perturbative HTL resummed gluon propagator~\cite{Guo:2020jvc}. Therefore, the regulation of the infrared divergence as well as the inclusion of the background field effect can be simultaneously realized with the theoretical method we used. This study is expected to give us some insight into the jet broadening in a ``semi-QGP" existing just above the critical temperature~\cite{Hidaka:2008dr}. We will report the corresponding results in a forthcoming paper~\cite{Ren:2026xfn}.

\section*{Acknowledgments}
The work is supported by the National Natural Science Foundation of China under Project No. 12465022 and by the Guangxi Science and Technology Program (No. Guike HJ2600640001).

\section*{Data availability}
The data that support the findings of this article are not publicly available. The data are available from the authors upon reasonable request.

\bibliographystyle{apsrev4-1}
\bibliography{paper}

@article{Shi:2018aeb,
    author = "Shi, Shao-wu and Jiang, Bing-feng and Hou, De-fu and Li, Jia-rong",
    title = "{The fluctuation energy exchange of a heavy quark in a collisional quark\textendash{}gluon plasma}",
    doi = "10.1016/j.nuclphysa.2018.09.048",
    journal = "Nucl. Phys. \textbf{A979}",
    pages = "265--275",
    year = "2018"
}

@article{Gyulassy:1990bh,
    author = "Gyulassy, Miklos and Plumer, Michael",
    editor = "Matthews, J. L. and Donnelly, T. W. and Farhi, E. H. and Osborne, L. S.",
    title = "{Jet quenching as a probe of dense matter}",
    reportNumber = "LBL-29569",
    doi = "10.1016/0375-9474(91)90173-4",
    journal = "Nucl. Phys. \textbf{A527}",
    pages = "641--644",
    year = "1991"
}

@misc{Bjorken:1982tu,
    author = {J. D. Bjorken},
    year = {1982},
    note = {{FERMILAB-PUB-82-059-THY}}
}

@article{Gyulassy:1993hr,
    author = "Gyulassy, Miklos and Wang, Xin-nian",
    title = "{Multiple collisions and induced gluon Bremsstrahlung in QCD}",
    eprint = "nucl-th/9306003",
    archivePrefix = "arXiv",
    reportNumber = "CU-TP-598, LBL-32682",
    doi = "10.1016/0550-3213(94)90079-5",
    journal = "Nucl. Phys. \textbf{B420}",
    pages = "583--614",
    year = "1994"
}

@article{Baier:1996kr,
    author = "Baier, R. and Dokshitzer, Yuri L. and Mueller, Alfred H. and Peigne, S. and Schiff, D.",
    title = "{Radiative energy loss of high-energy quarks and gluons in a finite volume quark - gluon plasma}",
    eprint = "hep-ph/9607355",
    archivePrefix = "arXiv",
    reportNumber = "BI-TP-96-21, CUTP-759, LPTHE-ORSAY-96-34",
    doi = "10.1016/S0550-3213(96)00553-6",
    journal = "Nucl. Phys. \textbf{B483}",
    pages = "291--320",
    year = "1997"
}

@article{Zakharov:1996fv,
    author = "Zakharov, B. G.",
    title = "{Fully quantum treatment of the Landau-Pomeranchuk-Migdal effect in QED and QCD}",
    eprint = "hep-ph/9607440",
    archivePrefix = "arXiv",
    doi = "10.1134/1.567126",
    journal = "JETP Lett.",
    volume = "63",
    pages = "952--957",
    year = "1996"
}

@article{Gyulassy:2000fs,
    author = "Gyulassy, M. and Levai, P. and Vitev, I.",
    title = "{NonAbelian energy loss at finite opacity}",
    eprint = "nucl-th/0005032",
    archivePrefix = "arXiv",
    reportNumber = "CU-TP-976",
    doi = "10.1103/PhysRevLett.85.5535",
    journal = "Phys. Rev. Lett.",
    volume = "85",
    pages = "5535--5538",
    year = "2000"
}

@article{Wiedemann:2000za,
    author = "Wiedemann, Urs Achim",
    title = "{Gluon radiation off hard quarks in a nuclear environment: Opacity expansion}",
    eprint = "hep-ph/0005129",
    archivePrefix = "arXiv",
    doi = "10.1016/S0550-3213(00)00457-0",
    journal = "Nucl. Phys. \textbf{B588}",
    pages = "303--344",
    year = "2000"
}

@article{Wang:2001ifa,
    author = "Wang, Xin-Nian and Guo, Xiao-feng",
    title = "{Multiple parton scattering in nuclei: Parton energy loss}",
    eprint = "hep-ph/0102230",
    archivePrefix = "arXiv",
    reportNumber = "LBNL-47155",
    doi = "10.1016/S0375-9474(01)01130-7",
    journal = "Nucl. Phys. \textbf{A696}",
    pages = "788--832",
    year = "2001"
}

@article{Arnold:2001ms,
    author = "Arnold, Peter Brockway and Moore, Guy D. and Yaffe, Laurence G.",
    title = "{Photon emission from quark gluon plasma: Complete leading order results}",
    eprint = "hep-ph/0111107",
    archivePrefix = "arXiv",
    reportNumber = "UW-PT-01-22",
    doi = "10.1088/1126-6708/2001/12/009",
    journal = "J. High Energy Phys.",
    volume = "12",
    year = "2001",
    pages = "009" 
}

@article{Arnold:2002ja,
    author = "Arnold, Peter Brockway and Moore, Guy D. and Yaffe, Laurence G.",
    title = "{Photon and gluon emission in relativistic plasmas}",
    eprint = "hep-ph/0204343",
    archivePrefix = "arXiv",
    reportNumber = "UW-PT-02-06",
    doi = "10.1088/1126-6708/2002/06/030",
    journal = "J. High Energy Phys.",
    volume = "06",
    pages = "030",
    year = "2002"
}

@article{Djordjevic:2003zk,
    author = "Djordjevic, Magdalena and Gyulassy, Miklos",
    title = "{Heavy quark radiative energy loss in QCD matter}",
    eprint = "nucl-th/0310076",
    archivePrefix = "arXiv",
    doi = "10.1016/j.nuclphysa.2003.12.020",
    journal = "Nucl. Phys. \textbf{A733}",
    pages = "265--298",
    year = "2004"
}

@article{Djordjevic:2006tw,
    author = "Djordjevic, Magdalena",
    title = "{Collisional energy loss in a finite size QCD matter}",
    eprint = "nucl-th/0603066",
    archivePrefix = "arXiv",
    doi = "10.1103/PhysRevC.74.064907",
    journal = "Phys. Rev. C",
    volume = "74",
    pages = "064907",
    year = "2006"
}

@article{Qin:2007rn,
    author = "Qin, Guang-You and Ruppert, Jorg and Gale, Charles and Jeon, Sangyong and Moore, Guy D. and Mustafa, Munshi G.",
    title = "{Radiative and collisional jet energy loss in the quark-gluon plasma at RHIC}",
    eprint = "0710.0605",
    archivePrefix = "arXiv",
    primaryClass = "hep-ph",
    doi = "10.1103/PhysRevLett.100.072301",
    journal = "Phys. Rev. Lett.",
    volume = "100",
    pages = "072301",
    year = "2008"
}

@article{Schenke:2009ik,
    author = "Schenke, Bjoern and Gale, Charles and Qin, Guang-You",
    title = "{The Evolving distribution of hard partons traversing a hot strongly interacting plasma}",
    eprint = "0901.3498",
    archivePrefix = "arXiv",
    primaryClass = "hep-ph",
    doi = "10.1103/PhysRevC.79.054908",
    journal = "Phys. Rev. C",
    volume = "79",
    pages = "054908",
    year = "2009"
}

@article{Dokshitzer:2001zm,
    author = "Dokshitzer, Yuri L. and Kharzeev, D. E.",
    title = "{Heavy quark colorimetry of QCD matter}",
    eprint = "hep-ph/0106202",
    archivePrefix = "arXiv",
    reportNumber = "LPT-ORSAY-01-58, BNL-NT-01-9",
    doi = "10.1016/S0370-2693(01)01130-3",
    journal = "Phys. Lett. B",
    volume = "519",
    pages = "199--206",
    year = "2001"
}

@article{Zhang:2003wk,
    author = "Zhang, Ben-Wei and Wang, Enke and Wang, Xin-Nian",
    title = "{Heavy quark energy loss in nuclear medium}",
    eprint = "nucl-th/0309040",
    archivePrefix = "arXiv",
    reportNumber = "LBNL-53791",
    doi = "10.1103/PhysRevLett.93.072301",
    journal = "Phys. Rev. Lett.",
    volume = "93",
    pages = "072301",
    year = "2004"
}

@article{Mustafa:2004dr,
    author = "Mustafa, Munshi G.",
    title = "{Energy loss of charm quarks in the quark-gluon plasma: Collisional versus radiative}",
    eprint = "hep-ph/0412402",
    archivePrefix = "arXiv",
    doi = "10.1103/PhysRevC.72.014905",
    journal = "Phys. Rev. C",
    volume = "72",
    pages = "014905",
    year = "2005"
}

@article{Wicks:2005gt,
    author = "Wicks, Simon and Horowitz, William and Djordjevic, Magdalena and Gyulassy, Miklos",
    title = "{Elastic, inelastic, and path length fluctuations in jet tomography}",
    eprint = "nucl-th/0512076",
    archivePrefix = "arXiv",
    doi = "10.1016/j.nuclphysa.2006.12.048",
    journal = "Nucl. Phys. \textbf{A784}",
    pages = "426--442",
    year = "2007"
}

@article{Schenke:2009gb,
    author = "Schenke, Bjoern and Gale, Charles and Jeon, Sangyong",
    title = "{MARTINI: An Event generator for relativistic heavy-ion collisions}",
    eprint = "0909.2037",
    archivePrefix = "arXiv",
    primaryClass = "hep-ph",
    doi = "10.1103/PhysRevC.80.054913",
    journal = "Phys. Rev. C",
    volume = "80",
    pages = "054913",
    year = "2009"
}

@article{Thoma:1990fm,
    author = "Thoma, Markus H. and Gyulassy, Miklos",
    title = "{Quark Damping and Energy Loss in the High Temperature {QCD}}",
    reportNumber = "LBL-29276",
    doi = "10.1016/S0550-3213(05)80031-8",
    journal = "Nucl. Phys. \textbf{B351}",
    pages = "491--506",
    year = "1991"
}

@article{Mrowczynski:1991da,
    author = "Mrowczynski, S.",
    title = "{Energy loss of a high-energy parton in the quark - gluon plasma}",
    doi = "10.1016/0370-2693(91)90188-V",
    journal = "Phys. Lett. B",
    volume = "269",
    pages = "383--388",
    year = "1991"
}

@article{Braaten:1991we,
    author = "Braaten, Eric and Thoma, Markus H.",
    title = "{Energy loss of a heavy quark in the quark - gluon plasma}",
    reportNumber = "LBL-30998, NUHEP-TH-91-14",
    doi = "10.1103/PhysRevD.44.R2625",
    journal = "Phys. Rev. D",
    volume = "44",
    number = "9",
    pages = "R2625",
    year = "1991"
}

@article{Romatschke:2003vc,
    author = "Romatschke, Paul and Strickland, Michael",
    title = "{Energy loss of a heavy fermion in an anisotropic QED plasma}",
    eprint = "hep-ph/0309093",
    archivePrefix = "arXiv",
    reportNumber = "TUW-03-09",
    doi = "10.1103/PhysRevD.69.065005",
    journal = "Phys. Rev. D",
    volume = "69",
    pages = "065005",
    year = "2004"
}

@article{Romatschke:2004au,
    author = "Romatschke, Paul and Strickland, Michael",
    title = "{Collisional energy loss of a heavy quark in an anisotropic quark-gluon plasma}",
    eprint = "hep-ph/0408275",
    archivePrefix = "arXiv",
    reportNumber = "TUW-04-19",
    doi = "10.1103/PhysRevD.71.125008",
    journal = "Phys. Rev. D",
    volume = "71",
    pages = "125008",
    year = "2005"
}

@article{Peigne:2007sd,
    author = "Peigne, Stephane and Peshier, Andre",
    title = "{Collisional Energy Loss of a Fast Muon in a Hot QED Plasma}",
    eprint = "0710.1266",
    archivePrefix = "arXiv",
    primaryClass = "hep-ph",
    doi = "10.1103/PhysRevD.77.014015",
    journal = "Phys. Rev. D",
    volume = "77",
    pages = "014015",
    year = "2008"
}

@article{Carrington:2003je,
    author = "Carrington, M. E. and Fugleberg, T. and Pickering, D. and Thoma, M. H.",
    title = "{Dielectric functions and dispersion relations of ultrarelativistic plasmas with collisions}",
    eprint = "hep-ph/0312103",
    archivePrefix = "arXiv",
    doi = "10.1139/p04-035",
    journal = "Can. J. Phys.",
    volume = "82",
    pages = "671--678",
    year = "2004"
}

@article{Zhao:2023mrz,
    author = "Zhao, Ruizhe and Qiu, Luhua and Guo, Yun and Strickland, Michael",
    title = "{Collective modes of a collisional anisotropic quark-gluon plasma}",
    eprint = "2306.12851",
    archivePrefix = "arXiv",
    primaryClass = "hep-ph",
    doi = "10.1103/PhysRevD.108.034023",
    journal = "Phys. Rev. D",
    volume = "108",
    number = "3",
    pages = "034023",
    year = "2023"
}

@article{Han:2017nfz,
    author = "Han, Cheng and Hou, De-fu and Jiang, Bing-feng and Li, Jia-rong",
    title = "{Jet energy loss in quark-gluon plasma: Kinetic theory with a Bhatnagar-Gross-Krook collisional kernel}",
    doi = "10.1140/epja/i2017-12400-9",
    journal = "Eur. Phys. J. A",
    volume = "53",
    number = "10",
    pages = "205",
    year = "2017"
}

@article{Jamal:2020emj,
    author = "Jamal, M. Yousuf and Mohanty, Bedangadas",
    title = "{Energy-loss of heavy quarks in the isotropic collisional hot QCD medium at a finite chemical potential}",
    eprint = "2002.09230",
    archivePrefix = "arXiv",
    primaryClass = "nucl-th",
    doi = "10.1140/epjp/s13360-021-01098-4",
    journal = "Eur. Phys. J. Plus",
    volume = "136",
    number = "1",
    pages = "130",
    year = "2021"
}

@article{YousufJamal:2019pen,
    author = "Yousuf Jamal, Mohammad and Chandra, Vinod",
    title = "{Energy loss of heavy quarks in the isotropic collisional hot QCD medium}",
    eprint = "1907.12033",
    archivePrefix = "arXiv",
    primaryClass = "nucl-th",
    doi = "10.1140/epjc/s10052-019-7278-2",
    journal = "Eur. Phys. J. C",
    volume = "79",
    number = "9",
    pages = "761",
    year = "2019"
}

@article{Schenke:2006xu,
    author = "Schenke, Bjoern and Strickland, Michael and Greiner, Carsten and Thoma, Markus H.",
    title = "{A Model of the effect of collisions on QCD plasma instabilities}",
    eprint = "hep-ph/0603029",
    archivePrefix = "arXiv",
    doi = "10.1103/PhysRevD.73.125004",
    journal = "Phys. Rev. D",
    volume = "73",
    pages = "125004",
    year = "2006"
}

@article{Hidaka:2008dr,
    author = "Hidaka, Yoshimasa and Pisarski, Robert D.",
    title = "{Suppression of the Shear Viscosity in a ''semi'' Quark Gluon Plasma}",
    eprint = "0803.0453",
    archivePrefix = "arXiv",
    primaryClass = "hep-ph",
    doi = "10.1103/PhysRevD.78.071501",
    journal = "Phys. Rev. D",
    volume = "78",
    pages = "071501",
    year = "2008"
}

@article{Lin:2013efa,
    author = "Lin, Shu and Pisarski, Robert D. and Skokov, Vladimir V.",
    title = "{Collisional energy loss above the critical temperature in QCD}",
    eprint = "1312.3340",
    archivePrefix = "arXiv",
    primaryClass = "hep-ph",
    doi = "10.1016/j.physletb.2014.01.043",
    journal = "Phys. Lett. B",
    volume = "730",
    pages = "236--242",
    year = "2014"
}

@article{Du:2024riq,
    author = "Du, Qianqian and Du, Mudong and Guo, Yun",
    title = "{Collisional energy loss of a heavy quark in a semiquark-gluon plasma}",
    eprint = "2402.18004",
    archivePrefix = "arXiv",
    primaryClass = "hep-ph",
    doi = "10.1103/PhysRevD.110.034011",
    journal = "Phys. Rev. D",
    volume = "110",
    number = "3",
    pages = "034011",
    year = "2024"
}

@article{Carignano:2021mrn,
    author = "Carignano, Stefano and Manuel, Cristina",
    title = "{Measuring chiral imbalance with collisional energy loss}",
    eprint = "2103.02491",
    archivePrefix = "arXiv",
    primaryClass = "hep-ph",
    doi = "10.1103/PhysRevD.103.116002",
    journal = "Phys. Rev. D",
    volume = "103",
    number = "11",
    pages = "116002",
    year = "2021"
}

@article{Guo:2024mgh,
    author = "Guo, Yun and Qiu, Luhua and Zhao, Ruizhe and Strickland, Michael",
    title = "{Energy loss of a heavy fermion in a collisional QED plasma}",
    eprint = "2403.06739",
    archivePrefix = "arXiv",
    primaryClass = "hep-ph",
    doi = "10.1103/PhysRevD.109.114025",
    journal = "Phys. Rev. D",
    volume = "109",
    number = "11",
    pages = "114025",
    year = "2024"
}

@article{Braaten:1991dd,
    author = "Braaten, Eric and Yuan, Tzu Chiang",
    title = "{Calculation of screening in a hot plasma}",
    reportNumber = "NUHEP-TH-91-01",
    doi = "10.1103/PhysRevLett.66.2183",
    journal = "Phys. Rev. Lett.",
    volume = "66",
    pages = "2183--2186",
    year = "1991"
}

@article{Peigne:2008nd,
    author = "Peigne, Stephane and Peshier, Andre",
    title = "{Collisional energy loss of a fast heavy quark in a quark-gluon plasma}",
    eprint = "0802.4364",
    archivePrefix = "arXiv",
    primaryClass = "hep-ph",
    doi = "10.1103/PhysRevD.77.114017",
    journal = "Phys. Rev. D",
    volume = "77",
    pages = "114017",
    year = "2008"
}

@article{Guo:2020jvc,
    author = "Guo, Yun and Kuang, Zhenpeng",
    title = "{Resummed gluon propagator and Debye screening effect in a holonomous plasma}",
    eprint = "2009.01516",
    archivePrefix = "arXiv",
    primaryClass = "hep-ph",
    doi = "10.1103/PhysRevD.104.014015",
    journal = "Phys. Rev. D",
    volume = "104",
    number = "1",
    pages = "014015",
    year = "2021"
}

@article{Dumitru:2010mj,
    author = "Dumitru, Adrian and Guo, Yun and Hidaka, Yoshimasa and Altes, Christiaan P. Korthals and Pisarski, Robert D.",
    title = "{How Wide is the Transition to Deconfinement?}",
    eprint = "1011.3820",
    archivePrefix = "arXiv",
    primaryClass = "hep-ph",
    reportNumber = "BNL-94342-2010-JA, KUNS-2310, NIKHEF-2010-040, RBRC-871",
    doi = "10.1103/PhysRevD.83.034022",
    journal = "Phys. Rev. D",
    volume = "83",
    pages = "034022",
    year = "2011"
}

@article{Dumitru:2012fw,
    author = "Dumitru, Adrian and Guo, Yun and Hidaka, Yoshimasa and Altes, Christiaan P. Korthals and Pisarski, Robert D.",
    title = "{Effective Matrix Model for Deconfinement in Pure Gauge Theories}",
    eprint = "1205.0137",
    archivePrefix = "arXiv",
    primaryClass = "hep-ph",
    reportNumber = "BNL-96946-2012-JA, NIKHEF-2012-002, RBRC-943, RIKEN-MP-40, RIKEN-QHP-20",
    doi = "10.1103/PhysRevD.86.105017",
    journal = "Phys. Rev. D",
    volume = "86",
    pages = "105017",
    year = "2012"
}

@article{Guo:2014zra,
    author = "Guo, Yun",
    title = "{Matrix Models for Deconfinement and Their Perturbative Corrections}",
    eprint = "1409.6539",
    archivePrefix = "arXiv",
    primaryClass = "hep-ph",
    doi = "10.1007/JHEP11(2014)111",
    journal = "J. High Energy Phys.",
    volume = "11",
    pages = "111",
    year = "2014"
}

@article{Faraday:2024gzx,
    author = "Faraday, Coleridge and Horowitz, W. A.",
    title = "{Collisional and radiative energy loss in small systems}",
    eprint = "2408.14426",
    archivePrefix = "arXiv",
    primaryClass = "nucl-th",
    doi = "10.1103/PhysRevC.111.054911",
    journal = "Phys. Rev. C",
    volume = "111",
    number = "5",
    pages = "054911",
    year = "2025"
}

@article{Djordjevic:2007at,
    author = "Djordjevic, Magdalena and Heinz, Ulrich",
    title = "{Radiative heavy quark energy loss in a dynamical QCD medium}",
    eprint = "0705.3439",
    archivePrefix = "arXiv",
    primaryClass = "nucl-th",
    doi = "10.1103/PhysRevC.77.024905",
    journal = "Phys. Rev. C",
    volume = "77",
    pages = "024905",
    year = "2008"
}

@article{Grishmanovskii:2025mnc,
    author = "Grishmanovskii, Ilia and Song, Taesoo and Greiner, Carsten and Bratkovskaya, Elena",
    title = "{Transport coefficients of heavy quarks by elastic and radiative scatterings in the strongly interacting quark-gluon plasma}",
    eprint = "2503.22311",
    archivePrefix = "arXiv",
    primaryClass = "hep-ph",
    doi = "10.1103/rb24-dg58",
    journal = "Phys. Rev. D",
    volume = "112",
    number = "1",
    pages = "014042",
    year = "2025"
}

@article{Faraday:2024zfj,
    author = "Faraday, Coleridge and Horowitz, W. A.",
    title = "{Heavy Flavour Energy Loss in Small and Large Systems}",
    eprint = "2409.14886",
    archivePrefix = "arXiv",
    primaryClass = "nucl-th",
    doi = "10.1051/epjconf/202531604001",
    journal = "EPJ Web Conf.",
    volume = "316",
    pages = "04001",
    year = "2025"
}

@article{Thoma:1993vs,
    author = "Thoma, M. H.",
    title = "{Parton interaction rates in the quark - gluon plasma}",
    eprint = "hep-ph/9308257",
    archivePrefix = "arXiv",
    reportNumber = "UGI-93-04",
    doi = "10.1103/PhysRevD.49.451",
    journal = "Phys. Rev. D",
    volume = "49",
    pages = "451--459",
    year = "1994"
}

@article{Kapusta:1991qp,
    author = "Kapusta, Joseph I. and Lichard, P. and Seibert, D.",
    title = "{High-energy photons from quark - gluon plasma versus hot hadronic gas}",
    doi = "10.1103/PhysRevD.47.4171",
    journal = "Phys. Rev. D",
    volume = "44",
    pages = "2774--2788",
    year = "1991",
    note = "[Erratum: Phys.Rev.D 47, 4171 (1993)]"
}

@article{Ren:2026xfn,
    author = "Ren, Haibo and Du, Qianqian and Guo, Yun",
    title = "{Suppression of the jet quenching parameter near the critical temperature}",
    eprint = "2601.11230",
    archivePrefix = "arXiv",
    primaryClass = "hep-ph",
    doi = "10.1103/zyg7-k87r",
    journal = "Phys. Rev. D",
    volume = "113",
    number = "5",
    pages = "054040",
    year = "2026"
}

\end{document}